\documentclass[bibyear]{aa} 

\usepackage{graphicx}
\usepackage{txfonts}
\begin{document}

\def\mincir{\raise -2.truept\hbox{\rlap{\hbox{$\sim$}}\raise5.truept \hbox{$<$}\ }}
\def\mincireq{\hbox{\raise0.5ex\hbox{$<\lower1.06ex\hbox{$\kern-1.07em{\sim}$}$}}}
\def\magcir{\raise-2.truept\hbox{\rlap{\hbox{$\sim$}}\raise5.truept \hbox{$>$}\ }}

\title{Diffuse non-thermal emission in the disks of the Magellanic Clouds}

\titlerunning{Relativistic particles in the Magellanic Clouds}


   \author{M. Persic
          \inst{1}
          \and
          Y. Rephaeli\inst{2}
          }

   \institute{
        INAF/Trieste Astronomical Observatory, via G.B.\,Tiepolo 11, I-34100 Trieste, Italy \\
        INFN-Trieste, via A.\,Valerio 2, I-34127 Trieste, Italy \\
              \email{massimo.persic@inaf.it}
        \and
        School of Physics \& Astronomy, Tel Aviv University, Tel Aviv 69978, Israel \\
        Center for Astrophysics and Space Sciences, University of California at San Diego, La Jolla, CA 92093, USA \\
             \email{yoelr@wise.tau.ac.il}
             }


 
  \abstract
{The Magellanic Clouds, two dwarf galaxy companions to the Milky Way, are among the {\it Fermi} Large Area 
Telescope (LAT) brightest $\gamma$-ray sources.}
{Comprehensive modeling of the non-thermal electromagnetic and neutrino emission 
in both Clouds.}
{We self-consistently model the radio and $\gamma$-ray spectral energy distribution from their disks based on 
recently published Murchison Widefield Array and {\it Fermi}/LAT data. All relevant radiative processes involving 
relativistic and thermal electrons (synchrotron, Compton scattering, and bremsstrahlung) and relativistic protons 
(neutral pion decay following interaction with thermal protons) are considered, using exact emission formulae. }
{Joint spectral analyses indicate that radio emission in the Clouds has both primary and secondary electron 
synchrotron and thermal bremsstrahlung origin, whereas $\gamma$\,rays originate mostly from $\pi^0$ decay with 
some contributions from relativistic bremsstrahlung and Compton scattering off starlight. The proton spectra in 
both galaxies are modeled as power laws in energy with similar spectral indices, $\sim$2.4, and energy densities, 
$\sim$1 eV cm$^{-3}$. The predicted 0.1--10\,GeV neutrino flux is too low for detection by current and upcoming 
experiments.}
{We confirm earlier suggestions of a largely hadronic origin of the $\gamma$-ray emission in both Magellanic Clouds.}


   \maketitle
%

\section{Introduction}

The spectral energy distributions (SEDs) of Cosmic Rays (CRs) outside their galactic accelerators are important 
for determining basic properties of CR populations and for assessing the impact of the particle interactions 
in the magnetized plasma in galactic disks and halos. Knowledge of these distributions is generally limited 
as it is usually based only on spectral radio observations. When, in addition, non-thermal (NT) X/$\gamma$-ray 
observations are available, reasonably detailed spectral modeling of the CR electron (CRe) and proton (CRp) 
distributions in star-forming environments can be very useful: Sampling the SED over these spectral regions 
yields important insight on the emitting CRe and possibly also CRp, whose interactions with the ambient plasma 
may dominate (via $\pi^0$ decay) high energy ($\magcir$100 MeV) emission.

The nearby Magellanic Clouds (MC), two irregular dwarf satellite galaxies of the Milky Way, may exemplify the 
level of spectral modeling currently feasible in a joint analysis of radio and $\gamma$-ray measurements. The 
Large MC (LMC) is located at a distance $d = 50$ kpc (Foreman et al. 2015); for the purpose of this study its 
structure is modeled as a cylinder with radius $r=3.5$ kpc and height $h=0.4$ kpc (Ackermann et al. 
2016). Likewise, for the Small MC (SMC) these quantities are $d = 60$\,kpc, $r=1.6$ kpc, and $h=4$ kpc (Abdo 
et al. 2010b). 

Both MCs have been detected with the {\it Fermi} Large Area Telescope (LAT) as extended sources (Abdo et al. 2010a,b). 
Based on recent data (Lopez et al. 2018), there is statistically significant emission along the SMC ''bar'' and 
''wing'' where there is active star formation. A spectro-spatial analysis of the LAT data suggests its integrated 
$\gamma$-ray emission to be mostly ($\geq$90\%) hadronic. 
Similar analyses of the LMC with multi-source models (Foreman et al. 2015; Ackermann et al. 2016; Tang et al. 2017) 
attempted to extract the spectral content of these sources in a statistically viable way (including details on the 
quality of the fits). The sources were found to be point-like (pulsars, supernova remnants, plerions, unidentified 
background AGNs) and extended. The latter included the large-scale disk denoted as 
\footnote
{
Emission components are labeled with ''E'' in Ackermann et al. (2016) and ''G'' in Tang et al. (2017).
}
E0 $\sim$ G1, and smaller-scale components denoted as E1+E3 $\sim$ G2 (the region west of 30 Doradus), E2 $\sim$ G3 (the 
LMC northern region), and E4 $\sim$ G4 (the LMC western region) -- all the latter possibly encompassing different sources 
with different spectra. These spectral/spatial analyses revealed a galaxy-scale component dominating the emission, with a 
spectral shape suggesting a lepto-hadronic (Foreman et al. 2015) or hadronic (Ackermann et al. 2016; Tang et al. 2017) 
origin.
\footnote{
As to the prominent 30 Dor massive-star--forming region, Ackermann et al. (2016) emphasize that it shines in GeV 
$\gamma$\,rays mainly because of the presence of PSR J0540-6919 and PSR J0537-6910.
} 

In this paper we focus on the emission in the MC disks in an attempt to determine mean values of their magnetic fields 
and CRe and CRp energy densities. In spite of their detailed LAT data analyses, the above-mentioned studies were not 
based on self-consistent modeling of the broadband radio/$\gamma$ SED of the two galaxies. SED modeling is important 
because a firm proof of the (mostly) pionic nature of the $\gamma$-ray emission should be based on a quantitative 
estimate of the CRe spectrum. Clearly, the latter is primarily determined from measurements of synchrotron radio 
emission -- which was not accounted for in those earlier works. In an attempt to clarify, and possibly remove, some 
uncertainties inherent in previous modeling work, here we re-assess key aspects of (average) conditions in the MC 
large-scale disks. Employing a one-zone model for the extended disk emission, we self-consistently carry out detailed 
calculations of the emission by CRe and CRp. The radio and $\gamma$-ray data used in our analyses are taken from the 
afore-mentioned papers Lopez et al. (2018; SMC), Tang et al. (2017; LMC), Ackermann et al. (2016; LMC), and For et al. 
(2018; SMC, LMC).

In Section 2 we briefly review the observations of extended NT emission from the MC disks. In section 3 we review the 
IR/optical radiation fields permeating the MCs. In Section 4 we describe calculations of the MC disk SEDs and perform 
fits to the data. Prospects of neutrino detection are discussed in Section 5, followed with a conclusion in Section 6.

\section{Observations of extended emission}

The MCs have been extensively observed over a wide range of radio/microwave, (soft) X-ray, and $\gamma$-ray bands. As 
mentioned above, point-source, and extended small and large scale emission has been detected from both galaxies. In 
this paper we focus on the extended large-scale disk emission of each galaxy because such emission traces the mean 
galactic properties of NT particles and magnetic fields. The spectral data sets used in our analysis are public (either 
tabulated or plotted) and are fully specified in Table\,1 (SMC) and Table\,2 (LMC). In this section we briefly review 
the observations most relevant to NT emission leaving out details that are discussed in the cited papers.
\bigskip

\noindent
$\bullet$ {\it Radio.}
In a multifrequency radio continuum study of the MCs, For et al. (2018) presented closely-sampled 76--227\,MHz 
Murchison Widefield Array data, supplemented by previous radio measurements at lower (for the LMC), and higher 
(for both MCs) frequencies. 
Measured fluxes include emission from MC (and background) point sources, which were 
estimated to contribute 11\% and 23\% of the measured emission of, respectively, the SMC 
and the LMC. Based on statistical analyses of a set of four fitting spectral models (power-law 
[1PL], curved PL, and double PL [2PL]) with either free and fixed high-$\nu$ index), they 
found that: 
\smallskip

\noindent
{\it (a)} for the SMC the best-fitting model is a 1PL (in the 76--8550 MHz band) with $\alpha($85.5\,MHz -- 8.55\,GHz$) 
= 0.82 \pm 0.03$ (consistent with Haynes et al. 1991); whereas 
\smallskip

\noindent
{\it (b)} for the LMC the preferred model is a 2PL (19.7--8550 MHz) with a low-$\nu$ (19.7--408 MHz) index 
$\alpha_0 = 0.66 \pm 0.08$ (consistent with Klein et al. 1989) and a (fixed) high-$\nu$ index $\alpha_1 = 0.1$ 
suggesting that synchrotron radiation and thermal free-free (ff) emission dominate at, respectively, low and high 
frequencies. 
\bigskip

\noindent
$\bullet$ {\it X-rays.}
The observed diffuse X-ray emission in the MCs has been determined to be of thermal origin (Wang et al. 1991: {\it 
Einstein Observatory}; Points et al. 2001: {\it R\"ontgen Satellit (ROSAT)} Position Sensitive Proportional Counters 
(PSPC); Nishiuchi et al. 1999, 2001: {\it Advanced Satellite for Cosmology and Astrophysics (ASCA)}); as such, it 
is not directly relevant to our SED analysis except for the estimated thermal plasma density, which is required to 
calculate the pionic emission (see below).
\bigskip

\noindent
$\bullet$ {\it $\gamma$-rays.}
Both MCs were detected at $>$100 MeV $\gamma$-rays as galaxy-scale extended sources whose emission is dominated by a 
diffuse component. The latter has been interpreted as mainly originating from CRp interacting with the interstellar 
gas. 
\smallskip

\noindent
{\it (a)} The SMC was first detected with 17 months of {\it Fermi}/LAT data (Abdo et al. 2010b) as a $\sim$3$^o$-size 
source, in which emission was not strongly correlated with prominent sites of star formation. More recently Lopez et al. 
(2018), based on 105 months of LAT Pass8 data, produced maps of the extended $>$2 GeV emission (no signal at $>$13 GeV) 
and found statistically significant emission along the galaxy-scale quietly--star-forming ''bar'' and ''wing''. Within 
a set of single-component spectral models -- i.e. PL, broken PL (with frozen spectral shape and free normalization, 
representing pulsars), and exponentially-cutoff PL (representing pionic emission) -- the latter provides the best fit 
to the total $\gamma$-ray spectrum, with only a marginal improvement of the fit if the broken PL is added. Lopez et al.'s 
analysis suggests that although pulsars may contribute $\leq$10\% at $>$100 MeV, the extended emission is mainly ($\geq$90\%) 
of pionic origin, with a $\gamma$-ray emissivity $\magcir$5 times smaller than the local (Galactic) one.
\smallskip

\noindent
{\it (b)} The LMC was marginally (>4.5\,$\sigma$) detected with the {\it Compton Gamma Ray Observatory}'s ({\it CGRO})
Energetic Gamma Ray Telescope (EGRET) (Sreekumar et al. 1992) and then confirmed (33\,$\sigma$) with 11 months of {\it 
Fermi}/LAT data (Abdo et al. 2010a). More recent studies have focused on the spatial and spectral modeling of the 
$\gamma$-ray surface brightness distribution; these are briefly reviewed below. 
\\
{\it (i)} The first LAT-based spectro-spatial analysis of the $\gamma$-ray surface brightness distribution was performed by 
Foreman et al. (2015) using 5.5 years of data, in the photon energy range 0.2$-$20 GeV. They modeled the CR distribution and 
$\gamma$-ray production based on observed maps of the LMC interstellar medium, star formation, radiation fields, and radio 
emission. The $\gamma$-ray spectrum was described by means of analytical fitting functions for the $\pi^0$-decay, bremsstrahlung, 
and inverse-Compton yields: these processes were estimated to account for, respectively, 50\%, 44\%, and 6\% of the total 
0.2$-$20 GeV emission. In particular they inferred the CRp spectral index to be $q_p = 2.4 \pm 0.2$ and the equipartition 
(with CRp) magnetic field to be $B_{\rm eq} = 2.8\mu$G. \\
{\it (ii)} A subsequent spectro-spatial analysis (Ackermann et al. 2016; A+16) of 6 years of LAT (P7REP) data in the 0.2--100 GeV 
band reported extended and point-like emissions. The (dominant) extended emission is in the form of a disk-scale component, 
denoted as E0 in their paper, and additional degree-scale emissions from several regions of enhanced star formation (including 
30 Doradus): if pionic, the E0 emission component implies a population of CRp with $\sim$1/3 the local Galactic density, 
whereas the superposed small-scale emissions imply local enhancements of the CRp density by factors of least 2-6. The spectrum 
of the E0 component (which does not include emission from the degree-scale emissions from several regions) 
shows some curvature (see their Fig.7-top): with no reference to the underlying emission mechanism, A+16 fitted the 
data with a tabulated function derived from the local (Galactic) gas emissivity spectrum (as a result of its interactions with 
CRp), an exponentially-cutoff PL, a broken PL, and a log-parabola, and concluded that the best-fitting analytical model was the 
log-parabola -- noting, however, that the similarity between the log-parabola and tabulated-function models (see their Fig.9) 
suggests a pionic nature of the large-scale disk emission. \\
{\it (iii)} More recently Tang et al. (2017; T+17) analyzed 8 years of LAT Pass8 data and deduced a deeper, more spectrally extended 
0.08--80 GeV spectrum of the large-scale disk component (identified as G1; this too did not include emission 
from other regions) largely overlaps with E0 of A+16. The G1 shape, now extending down to 60 MeV, was determined 
to be best described as a $\pi^0$-decay hump from a CRp spectrum harder than that in our Galaxy -- although they could not rule out 
CRe relativistic bremsstrahlung completely. \\[0.2cm]
In this analysis we use the T+17 G1 data as our reference set for the diffuse large-scale LMC disk emission, and 
the earlier A+16 E0 data as an auxiliary set for a consistency check.

\begin{table*}
\caption[] {SMC radio and $\gamma$-ray data.}
\centering 
\begin{tabular}{ l  l  l  l  l  l  l}
\hline
\hline
\noalign{\smallskip}
Frequency    &        Energy Flux      & Reference             &  & Frequency                                &   Energy Flux           & Reference              \\
log($\nu$/Hz)& $10^{-12}$erg/(cm$^2$s) &                       &  &log($\nu$/Hz)                             & $10^{-12}$erg/(cm$^2$s) &            \\
\noalign{\smallskip}
\hline
\noalign{\smallskip}
7.932        &     $0.393 \pm 0.171$   & Mills (1955, 1959)    &  &   8.340                                  &  $0.481 \pm 0.125$      & For et al. (2018)      \\
7.881        &     $0.307 \pm 0.080$   & For et al. (2018)     &  &   8.356                                  &  $0.489 \pm 0.127$      & For et al. (2018)      \\
7.924        &     $0.245 \pm 0.064$   & For et al. (2018)     &  &   8.611                                  &  $0.543 \pm 0.041$      & Loiseau et al. (1987)  \\
7.964        &     $0.230 \pm 0.060$   & For et al. (2018)     &  &   9.146                                  &  $0.588 \pm 0.084$      & Loiseau et al. (1987)  \\
7.996        &     $0.260 \pm 0.068$   & For et al. (2018)     &  &   9.146                                  &  $0.486 \pm 0.028$      & For et al. (2018)      \\
8.029        &     $0.379 \pm 0.099$   & For et al. (2018)     &  &   9.362                                  &  $0.713 \pm 0.138$      & Mountfort et al. (1987)\\
8.061        &     $0.311 \pm 0.081$   & For et al. (2018)     &  &   9.362                                  &  $0.623 \pm 0.161$      & For et al. (2018)      \\
8.090        &     $0.316 \pm 0.082$   & For et al. (2018)     &  &   9.389                                  &  $0.637 \pm 0.073$      & Haynes et al. (1991)   \\
8.114        &     $0.317 \pm 0.083$   & For et al. (2018)     &  &   9.677                                  &  $0.902 \pm 0.190$      & Haynes et al. (1991)   \\
8.155        &     $0.464 \pm 0.121$   & For et al. (2018)     &  &   9.932                                  &  $1.282 \pm 0.342$      & Haynes et al. (1991)   \\
8.176        &     $0.387 \pm 0.101$   & For et al. (2018)     &  & $22.860^{\scriptscriptstyle +0.125}_{\scriptscriptstyle -0.167}$ &  $7.047 \pm 0.860$   & Lopez et al. 2018      \\
8.199        &     $0.417 \pm 0.109$   & For et al. (2018)     &  & $23.161^{\scriptscriptstyle +0.125}_{\scriptscriptstyle -0.167}$ &  $7.998 \pm 0.853$   & Lopez et al. 2018      \\
8.220        &     $0.360 \pm 0.094$   & For et al. (2018)     &  & $23.462^{\scriptscriptstyle +0.125}_{\scriptscriptstyle -0.167}$ &  $7.194 \pm 0.658$   & Lopez et al. 2018      \\
8.241        &     $0.620 \pm 0.161$   & For et al. (2018)     &  & $23.781^{\scriptscriptstyle +0.125}_{\scriptscriptstyle -0.167}$ &  $5.284 \pm 0.563$   & Lopez et al. 2018      \\
8.258        &     $0.512 \pm 0.133$   & For et al. (2018)     &  & $24.055^{\scriptscriptstyle +0.125}_{\scriptscriptstyle -0.167}$ &  $5.598 \pm 0.683$   & Lopez et al. 2018      \\
8.276        &     $0.473 \pm 0.123$   & For et al. (2018)     &  & $24.383^{\scriptscriptstyle +0.125}_{\scriptscriptstyle -0.167}$ &  $3.357 \pm 0.754$   & Lopez et al. 2018      \\
8.294        &     $0.486 \pm 0.127$   & For et al. (2018)     &  & $24.684^{\scriptscriptstyle +0.125}_{\scriptscriptstyle -0.167}$ &  $<2.972$            & Lopez et al. 2018      \\
8.310        &     $0.594 \pm 0.155$   & For et al. (2018)     &  & $24.985^{\scriptscriptstyle +0.125}_{\scriptscriptstyle -0.167}$ &  $<0.879$            & Lopez et al. 2018      \\
8.326        &     $0.489 \pm 0.127$   & For et al. (2018)  \\
\noalign{\smallskip}
\hline\end{tabular}
\end{table*}

\begin{table*}
\caption[] {LMC radio and $\gamma$-ray data.}
\centering 
\begin{tabular}{ l  l  l  l  l  l  l}
\hline
\hline
\noalign{\smallskip}
Frequency      &             Energy Flux     & Reference              &   & Frequency           &   Energy Flux         & Reference               \\
log($\nu$/Hz)  &      $10^{-12}$erg/(cm$^2$s)&                        &   & log($\nu$/Hz)       &$10^{-12}$erg/(cm$^2$s)&                         \\
\noalign{\smallskip}
\hline
\noalign{\smallskip}
7.294     &     $1.038 \pm 0.208$   & Shain (1959)                    &   & 9.389                &  $9.555 \pm 0.490$    & Haynes et al. (1991)    \\
7.653     &     $1.349 \pm 0.203$   & Alvarez et al. (1987)           &   & 9.677                &  $17.243 \pm 1.425$   & Haynes et al. (1991)    \\
7.881     &     $1.410 \pm 0.240$   & For et al. (2018)               &   & 9.677                &  $14.250 \pm 1.900$   & For et al. (2018)       \\
7.924     &     $1.492 \pm 0.254$   & For et al. (2018)               &   & 9.932                &  $23.085 \pm 2.993$   & Haynes et al. (1991)    \\
7.932     &     $3.154 \pm 0.342$   & Mills (1959)                    &   &                      & $=========$           &                         \\
7.964     &     $1.449 \pm 0.246$   & For et al. (2018)               &   &                      & $10^{-11}$erg/(cm$^2$s) &                       \\
7.986     &     $2.748 \pm 0.581$   & Mills (1959)                    &   &                      & --------------------- &                         \\
7.996     &     $1.437 \pm 0.245$   & For et al. (2018)               &   & $22.2946 \pm 0.1336$ &   $0.93 \pm 0.31$     &   Tang et al. (2017)    \\
8.029     &     $1.956 \pm 0.333$   & For et al. (2018)               &   & $22.5635 \pm 0.1353$ &   $1.64 \pm 0.33$     &   Tang et al. (2017)    \\
8.061     &     $1.872 \pm 0.318$   & For et al. (2018)               &   & $22.8325 \pm 0.1336$ &   $1.91 \pm 0.19$     &   Tang et al. (2017)    \\
8.090     &     $2.021 \pm 0.344$   & For et al. (2018)               &   & $23.1004 \pm 0.1343$ &   $2.08 \pm 0.13$     &   Tang et al. (2017)    \\
8.114     &     $2.043 \pm 0.347$   & For et al. (2018)               &   & $23.3691 \pm 0.1344$ &   $2.17 \pm 0.15$     &   Tang et al. (2017)    \\
8.155     &     $2.379 \pm 0.405$   & For et al. (2018)               &   & $23.6376 \pm 0.1342$ &   $1.57 \pm 0.15$     &   Tang et al. (2017)    \\
7.653     &     $1.349 \pm 0.203$   & Alvarez et al. (1987)           &   & $23.9061 \pm 0.1343$ &   $1.44 \pm 0.17$     &   Tang et al. (2017)    \\
8.176     &     $2.175 \pm 0.370$   & For et al. (2018)               &   & $24.1747 \pm 0.1342$ &   $1.16 \pm 0.21$     &   Tang et al. (2017)    \\
8.199     &     $2.134 \pm 0.363$   & For et al. (2018)               &   & $24.4432 \pm 0.1343$ &   $0.52 \pm 0.24$     &   Tang et al. (2017)    \\
8.199     &     $2.743 \pm 0.774$   & Mills (1959)                    &   & $24.7117 \pm 0.1342$ &   $1.09 \pm 0.31$     &   Tang et al. (2017)    \\
8.220     &     $1.999 \pm 0.340$   & For et al. (2018)               &   & $24.9801 \pm 0.1342$ &       $<0.73$         &   Tang et al. (2017)    \\
8.241     &     $2.324 \pm 0.397$   & For et al. (2018)               &   & $25.2486 \pm 0.1342$ &   $0.63 \pm 0.36$     &   Tang et al. (2017)    \\
8.258     &     $2.257 \pm 0.384$   & For et al. (2018)               &   & $22.781 \pm 0.100$   &  $3.524 \pm 0.912$   & Ackermann et al. (2016)  \\
8.276     &     $2.313 \pm 0.393$   & For et al. (2018)               &   & $22.985 \pm 0.100$   &  $3.357 \pm 0.349$   & Ackermann et al. (2016)  \\
8.294     &     $2.186 \pm 0.372$   & For et al. (2018)               &   & $23.182 \pm 0.100$   &  $3.041 \pm 0.309$   & Ackermann et al. (2016)  \\
8.310     &     $2.520 \pm 0.429$   & For et al. (2018)               &   & $23.383 \pm 0.100$   &  $2.958 \pm 0.256$   & Ackermann et al. (2016)  \\
8.326     &     $2.378 \pm 0.404$    & For et al. (2018)              &   & $23.587 \pm 0.100$   &  $2.685 \pm 0.232$   & Ackermann et al. (2016)  \\
8.340     &     $2.261 \pm 0.385$    & For et al. (2018)              &   & $23.781 \pm 0.100$   &  $1.811 \pm 0.254$   & Ackermann et al. (2016)  \\
8.356     &     $2.315 \pm 0.394$    & For et al. (2018)              &   & $23.985 \pm 0.100$   &  $0.859 \pm 0.240$   & Ackermann et al. (2016)  \\
8.611     &     $3.774 \pm 0.122$    & Klein et al. (1989)            &   & $24.161 \pm 0.100$   &  $1.167 \pm 0.326$   & Ackermann et al. (2016)  \\
9.146     &     $5.367 \pm 0.420$    & For et al. (2018)              &   & $24.383 \pm 0.100$   &  $0.980 \pm 0.315$   & Ackermann et al. (2016)  \\
9.146     &     $7.406 \pm 0.420$    & Klein et al. (1989)            &   & $24.587 \pm 0.100$   &      $<0.706$        & Ackermann et al. (2016)  \\
9.362     &     $9.476 \pm 1.150$    & Mountfort et al. (1987)        &   & $25.183 \pm 0.504$   &      $<1.119$        & Ackermann et al. (2016)  \\
\noalign{\smallskip}
\hline\end{tabular}
\end{table*}

\section{Radiation fields}

A reasonably precise determination of the ambient radiation field is needed for predicting the level of $\gamma$-ray emission from 
Compton scattering of the radio-emitting electrons (and positrons). The total radiation field includes cosmic (background) and local 
(foreground) components.

Relevant cosmic radiation fields include the Cosmic Microwave Background (CMB) and the Extragalactic Background Light (EBL). The CMB is 
a pure Planckian described by a temperature $T_{\rm CMB}=2.735\,(1+z)$ K and energy density $u_{\rm CMB} = 0.25\,(1+z)^4$ eV cm$^{-3}$. 
The EBL originates from direct and dust-reprocessed starlight integrated over the star formation history the Universe. It shows two peaks, 
corresponding respectively to the Cosmic Infrared Background (CIB, at $\sim$100\,$\mu$m) that originates from dust-reprocessed starlight 
integrated over the star formation history of galaxies, and the Cosmic Optical Background (COB, at $\sim$1\,$\mu$m) that originates from 
direct starlight integrated over all stars that formed (e.g. Cooray 
2016). The two peaks are described as diluted Planckians, characterized by a temperature $T$ and a dilution factor $C_{\rm dil}$. The 
latter is the ratio of the actual energy density, $u$, to the energy density of an undiluted blackbody at the same temperature $T$, 
i.e. $u = C_{\rm dil}\, a T^4$, where $a$ is the Stefan-Boltzmann constant. The dilution factors of the cosmic fields are, $C_{\rm CMB} 
= 1$, $C_{\rm CIB} = 10^{-5.629}$, and $C_{\rm COB} = 10^{-13.726}$. A recent updated EBL model, based on accurate galaxy counts in 
several spectral bands, is due to Franceschini \& Rodighiero (2017); locally ($z=0$) it can be numerically approximated as a combination 
of diluted Planckians, 
\begin{eqnarray}
\lefteqn{
n_{\rm EBL}(\epsilon) ~=~ \sum_{j=1}^8 A_j \,\frac{8 \pi}{h^3c^3} \, \frac{\epsilon^2}{e^{\epsilon/k_B T_j}-1} \hspace{0.5cm}  
{\rm cm^{-3}~ erg^{-1}} }
\label{eq:EBL}
\end{eqnarray}
with: 
$A_1=10^{-5.629}$, $T_1=29$\,K; 
$A_2=10^{-8.522}$, $T_2=96.7$\,K; 
$A_3=10^{-10.249}$, $T_3=223$\,K; 
$A_4=10^{-12.027}$, $T_4=580$\,K;  
$A_5=10^{-13.726}$, $T_5=2900$\,K; 
$A_6=10^{-15.027}$, $T_6=4350$\,K; 
$A_7=10^{-16.404}$, $T_7=5800$\,K;
$A_8=10^{-17.027}$, $T_8=11600$\,K. 

Local radiation fields in the MCs arise from their intrinsic stellar populations and (given its close proximity) also from the 
Milky Way; we refer to this as the Galactic Foregound Light (GFL). Similar to the EBL by shape and origin, the GFL is dominated 
by two thermal humps, IR and optical. The corresponding energy densities (in eV cm$^{-3}$) are: {\it (i)} LMC: $u_{\rm IR} = 0.12$ 
and $u_{\rm opt} = 0.20$, estimated from Foreman et al. 2015 ($u_{\rm IR+opt} = 0.32$) and the IR/opt SED (from NED
\footnote{
The NASA/IPAC Extragalactic Database (NED) is operated by the Jet Propulsion Laboratory (JPL), Caltech, 
under contract with the National Aeronautic and Space Administration (NASA).
}
); {\it (ii)} 
SMC: $u_{\rm IR} = 0.026$, $u_{\rm opt} = 0.062$, scaling down the LMC values taking (from the SED, cf. NED) the optical and IR 
peaks to be factors of 0.215 and 0.144, respectively, of the corresponding LMC values.

%
\begin{figure}
\vspace{14.0cm}
\includegraphics{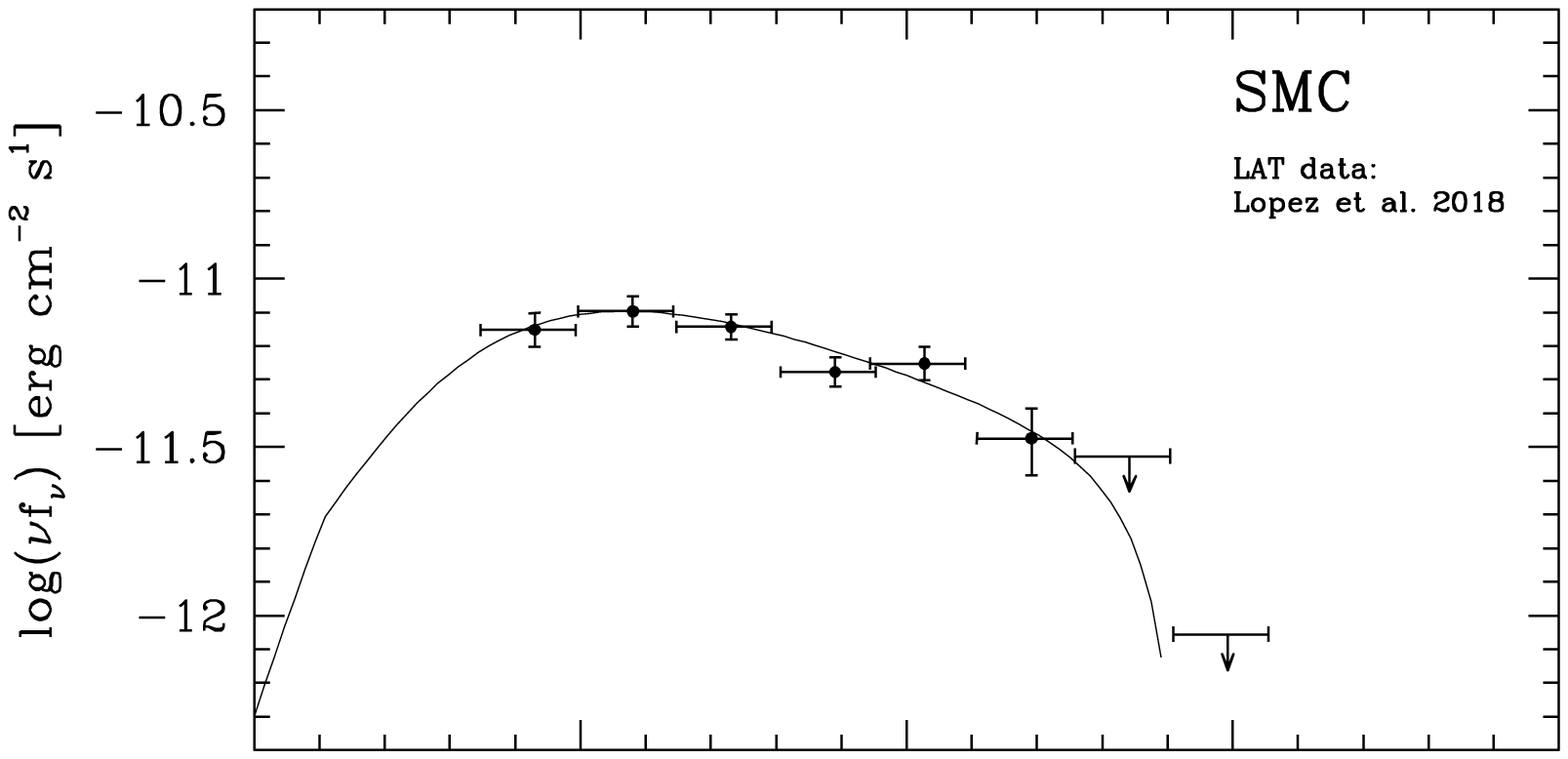}
\includegraphics{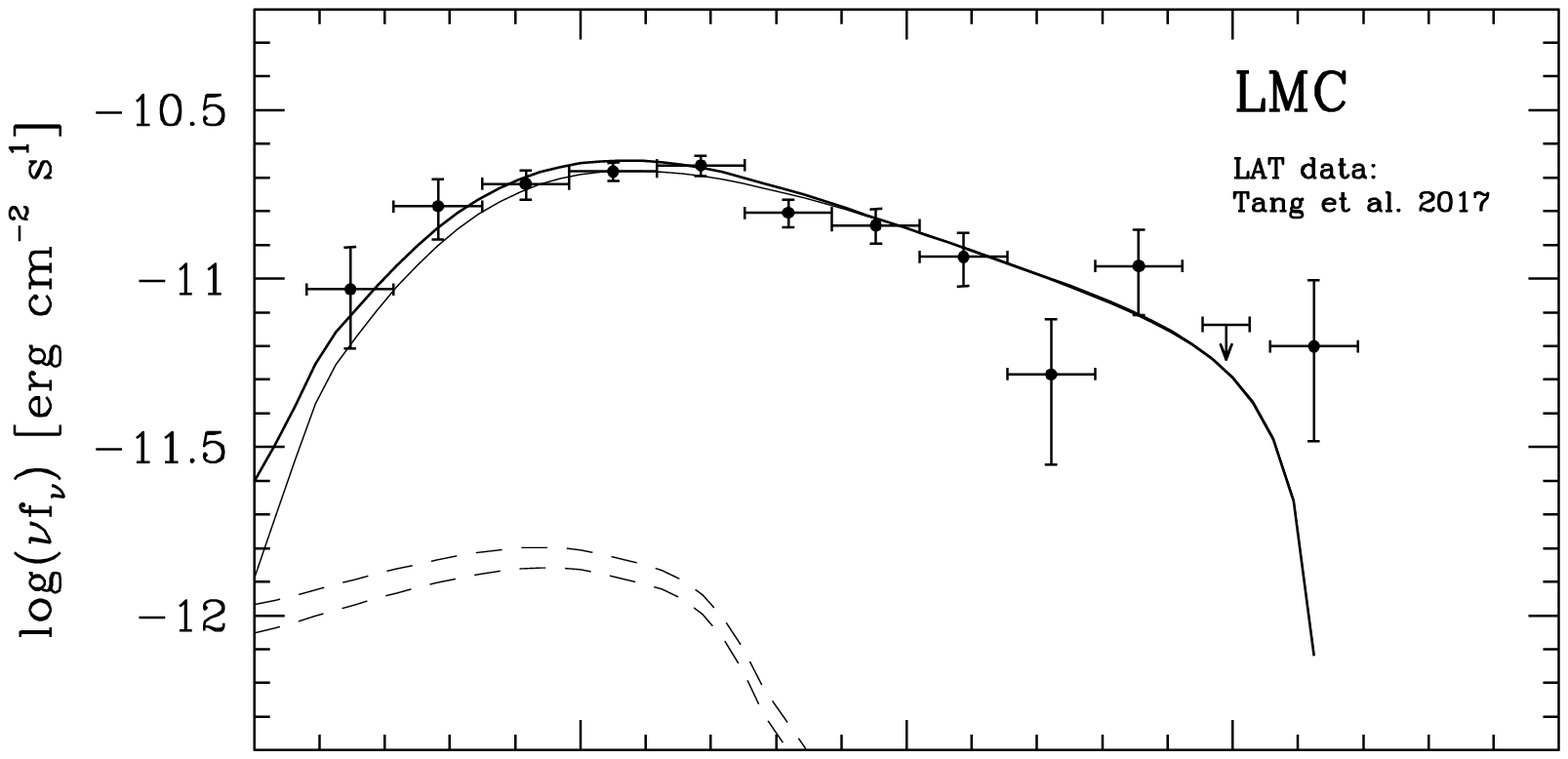}
\includegraphics{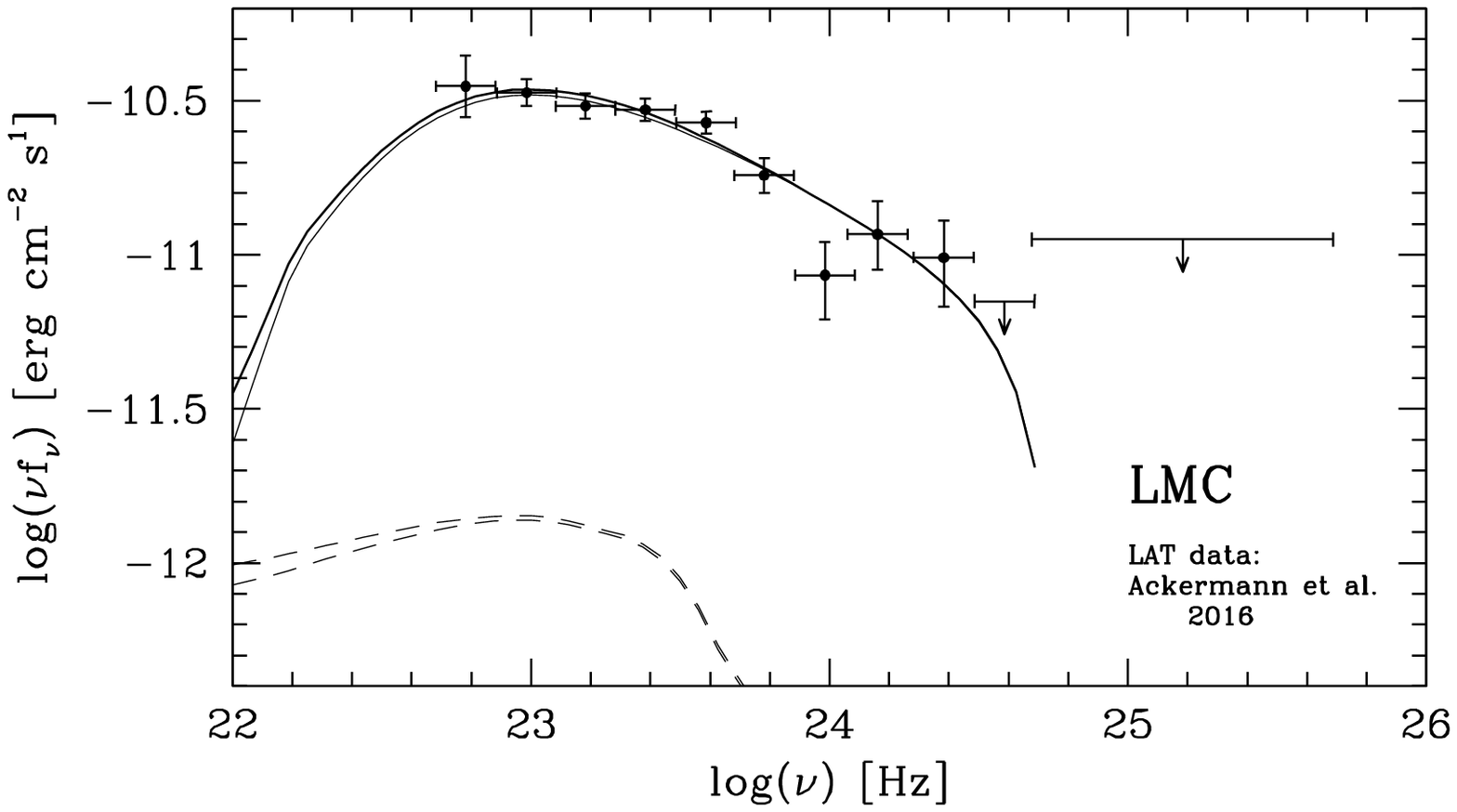}
\caption{
The MC $\gamma$-ray spectra. Symbols are as in Fig.\ref{fig:SMC_SED}.
}
\label{fig:gamma}
\end{figure}
%

%
\begin{figure*}
\vspace{4.7cm}
\includegraphics{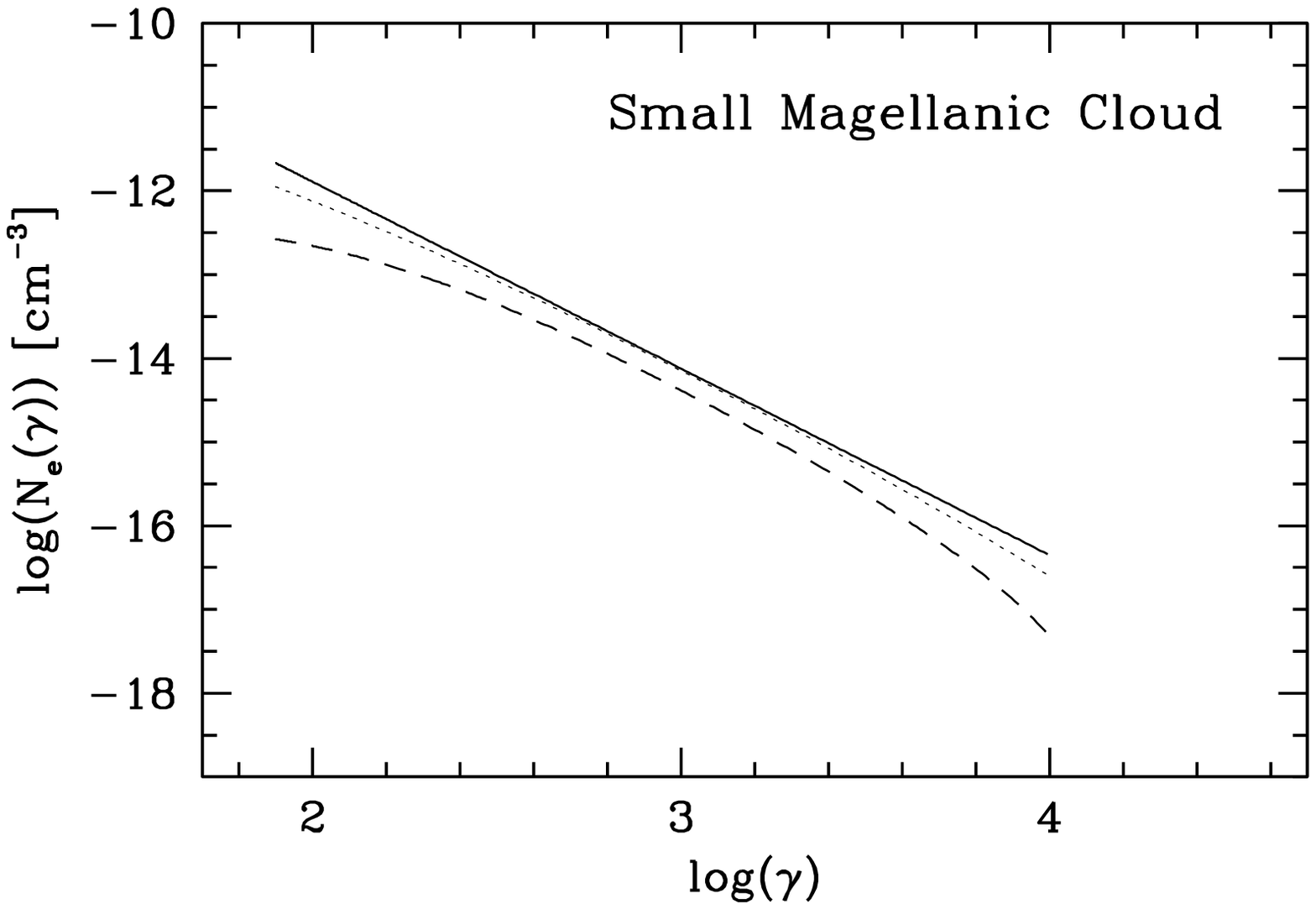}
\includegraphics{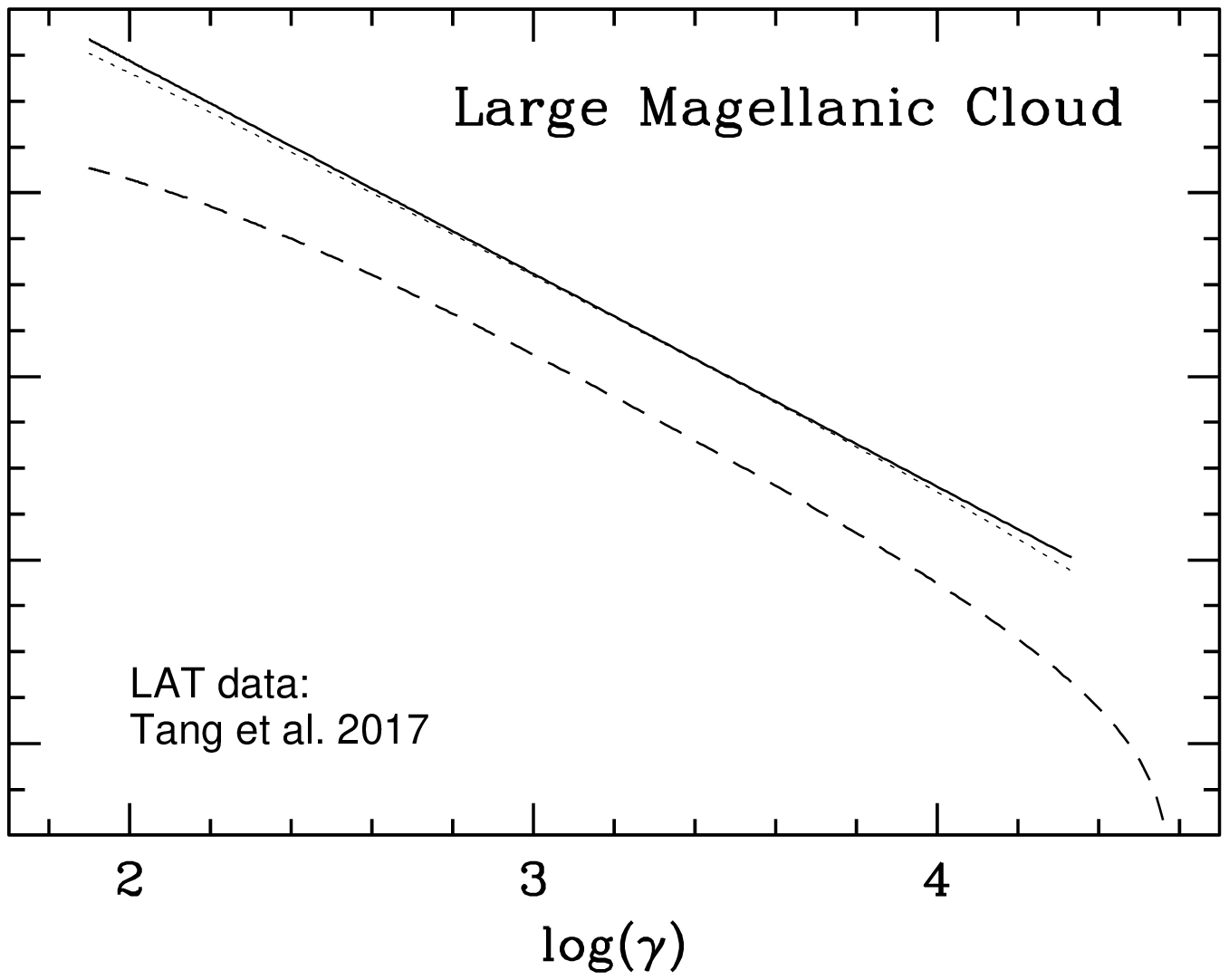}
\includegraphics{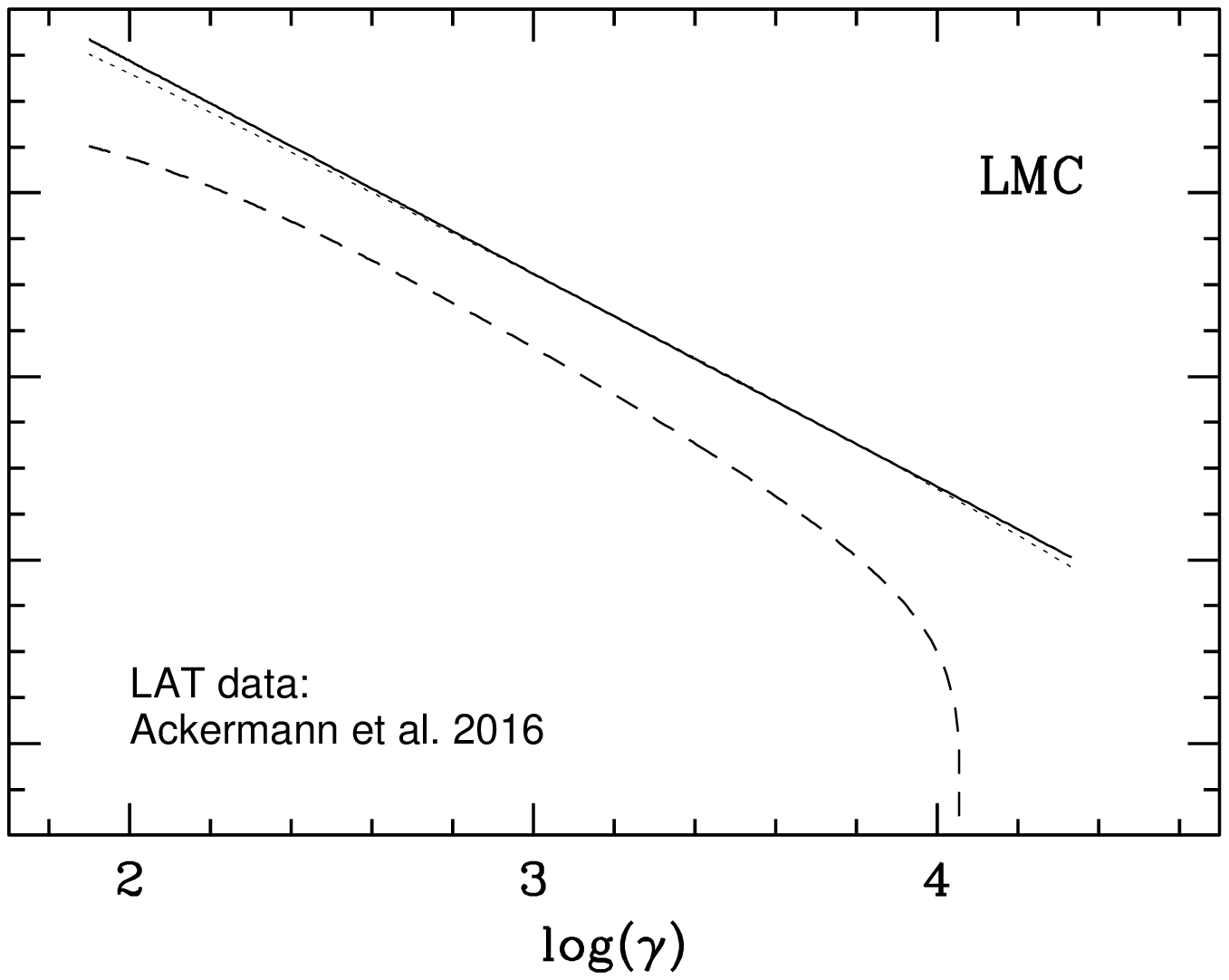}
\caption{
CRe spectra in the Magellanic Clouds. Curves denote secondary spectra (dashed), primary spectra (dotted), 
and the latter's local PL representations (solid). In each panel, the normalization of the primary-CRe 
spectrum to its PL counterpart is chosen such that they nearly overlap. The primary-CRe injection indices, 
$q_i$, are 2.28 (SMC) and 2.23 (LMC). The two LMC CRe spectra panels refer, respectively, to the 
T+17 and A+16 {\it Fermi}/LAT datasets, as secondary spectra are derived 
from the pionic fits to the $\gamma$-ray data. 
}
\label{fig:MC_CRe_spectra}
\end{figure*}
%

%
\begin{figure*}
\vspace{6.2cm}
\includegraphics{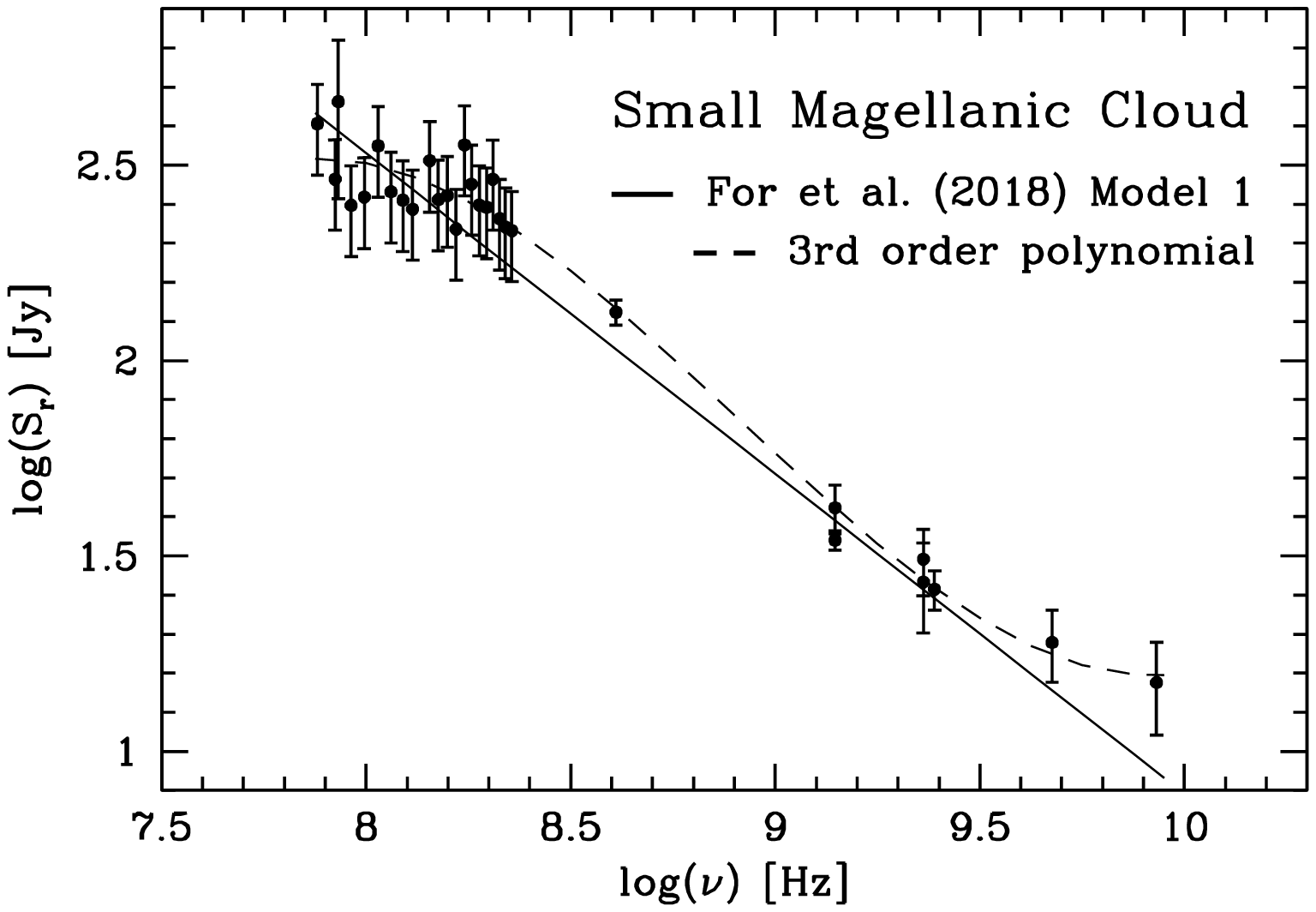}
\includegraphics{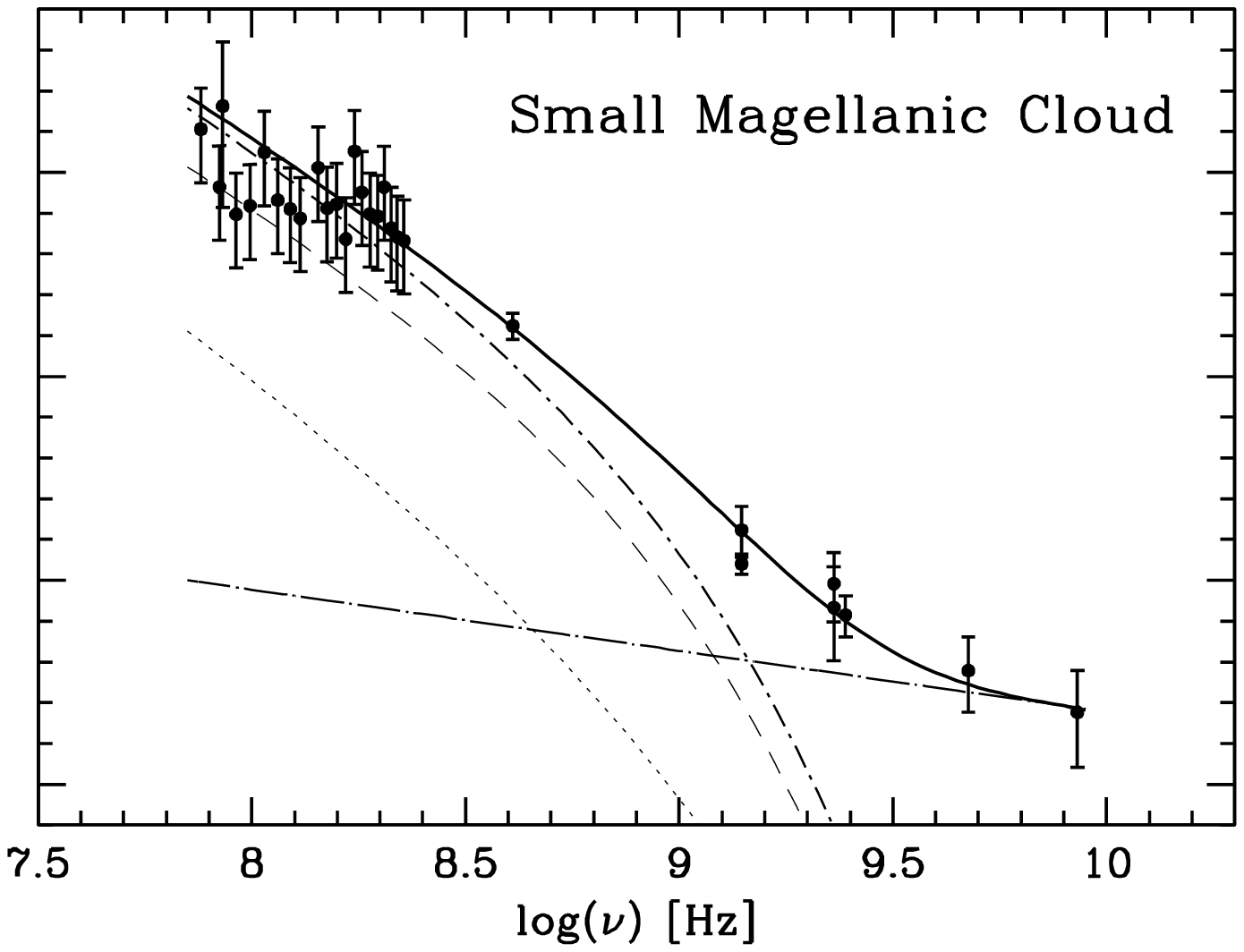}
\caption{ 
SMC radio spectrum: data points (dots) are from Table 3 of For et al. (2018). 
{\it Left:} the "preferred" 1PL model\,1 of For et al. (2018) is shown as a solid line, 
and the best fitting 3rd-order polynomial described in Table 3 is shown as a dashed curve.
{\it Right:} the two-component model (solid curve) that mimicks the polynomial includes 
synchrotron radiation (dotted/short-dashed curve; dashed, dotted curves denote primary, 
secondary contributions) and thermal-ff emission (dotted/long-dashed curve). 
}
\label{fig:SMC_radio}
\end{figure*}
%

%
\begin{figure}
\vspace{5.4cm}
\includegraphics{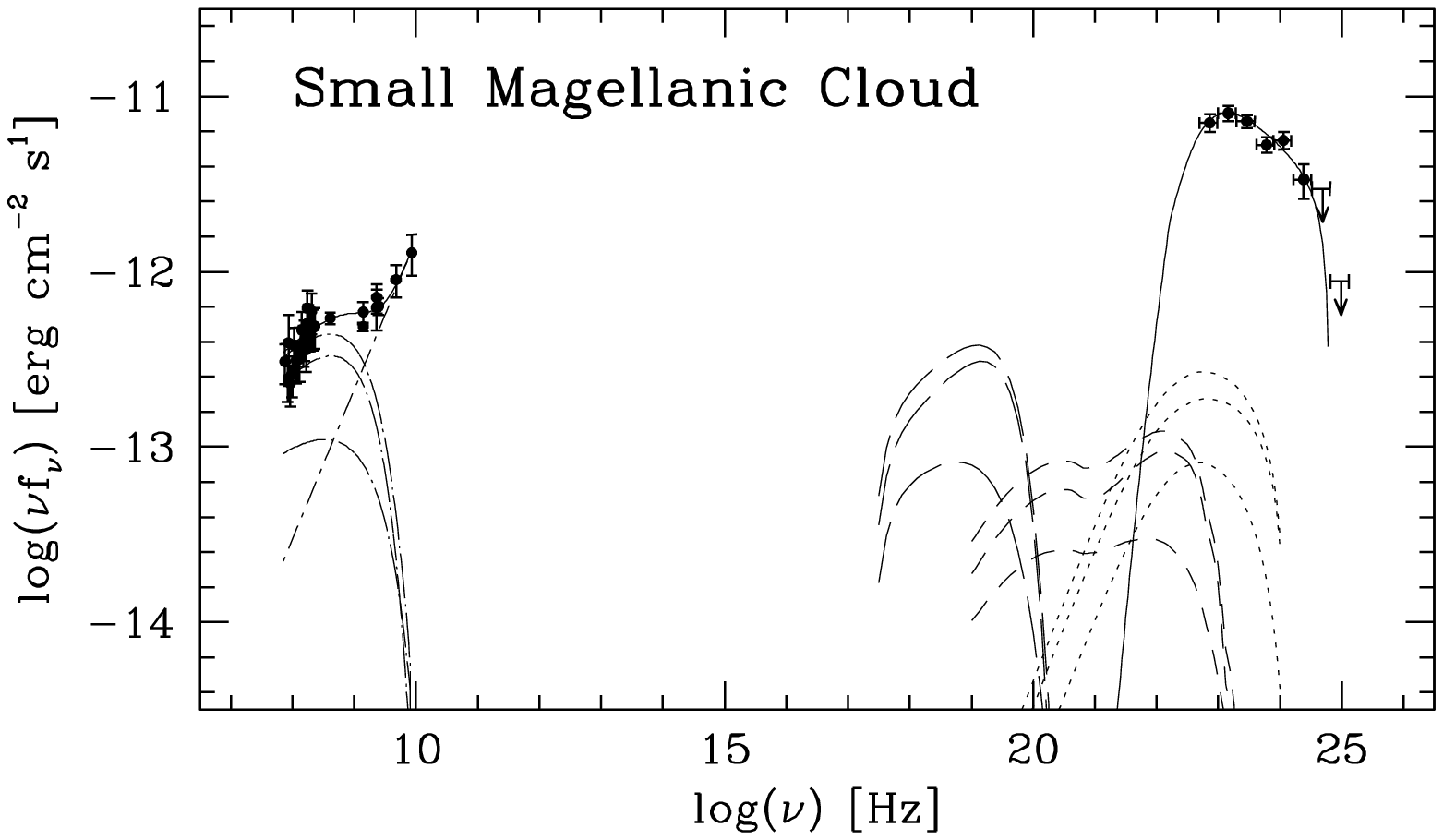}
\caption{ 
The broad-band (76\,MHz -- 53.7\,GeV) NT SED of the SMC disk. Data points (Table 1) are shown by dots, 
model predictions by curves. Emission components are plotted by the following line types: 
synchrotron, dotted/long-dashed; 
thermal-ff, dotted/short-dashed; 
total radio, solid; Compton/CMB, long-dashed; Compton/starlight, short-dashed; 
NT bremsstrahlung, dotted; pionic, solid. For each type of NT leptonic yield secondary, primary, 
and total emissions are denoted as curves of progressively higher flux. 
}
\label{fig:SMC_SED}
\end{figure}
%

\section{SED models}

Our main objective in this study is to determine CR electron and proton spectra in the MC disks 
by spectral modeling of NT emission in all relevant energy range accessible to observations. The 
particle, gas density, magnetic and radiation field distributions clearly 
vary significantly across the disk, obviating the need for a spectro-spatial treatment. Indeed, 
a modeling approach based on a solution to the diffusion-advection equation has been applied 
in the study of the Galaxy and several nearby galaxies. However, even for (just) a 
diffusion-based treatment to be feasible and meaningful, the spatial profiles of key 
quantities, such as the particle acceleration sources, gas density, magnetic field, and 
generally also the diffusion coefficient have to be specified. Given the generally very 
limited observational basis for reliably determining these profiles, a parameter-intensive 
spectro-spatial modeling approach is rarely warranted. 
An example is the very approximate diffusion-based approach adopted 
in modeling NT emission in the disks and halos of the star-forming galaxies NGC4631 and 
NGC4666 (Rephaeli \& Sadeh 2019) for which reasonably detailed radio spectra and spatial 
profiles are available; even so, the results of these analyses are not conclusive given the 
substantial uncertainty in the values of key parameters.

Radio emission in the MC disks is electron synchrotron in a disordered magnetic field whose mean 
value $B$ is taken to be spatially uniform, and thermal-ff from a warm ionized plasma. A significant 
CRp component could yield additional radio emission by secondary $e^{\pm}$ produced by $\pi^{\pm}$ 
decays, and $\gamma$-ray emission from $\pi^{0}$ decay. In addition, $\gamma$-ray emission is 
produced by CRe scattering off photons of local and background radiation fields. The calculations of 
the emissivities from all these processes are well known and standard. 

We assume the CR spectral distributions to be time-independent and locally isotropic with spectral PL form. The primary CRe 
spectral density (per cm$^3$ and per unit of the electron Lorentz factor $\gamma$) is, $N_e(\gamma) = N_{e,0} \, \gamma^{-q_e}$ 
for $\gamma_{min} < \gamma < \gamma_{max}$ with $\gamma_{min}>>1$. As discussed later, this spectrum proves to be a good 
approximation to the actual primary steady-state spectrum in the relevant electron energy range. The secondary CRe spectrum can 
be analytically approximated as a smoothed-2PL, exponentially cutoff at high energies (see below). The assumed CRp spectrum (in 
units of cm$^{-3}$ GeV$^{-1}$) as a function of $E_p$/GeV is, $N_p(E_p) = N_{p,0} \, E_p^{-q_p}$ for $m_p < E < E_p^{max}$.

\subsection{SMC} 

The emission spectrum from $\pi^{0}$-decay has more constraining power than a generic PL; thus, our modeling procedure begins 
with fitting a pionic emission profile to the $\gamma$-ray data with free normalization, slope, and high-energy cutoff. We then 
use the deduced (essentially, fully-determined) secondary-CRe spectrum together with an assumed primary-CRe spectrum to calculate 
the combined synchrotron emission using an observationally estimated value of $B$. The fit to the radio data includes, at high 
frequencies, also a thermal bremsstrahlung component computed with previously determined values of the gas density and temperature. 
Finally, the full Compton X/$\gamma$-ray yield is determined.

The $\pi^0$-decay $\gamma$-ray flux is computed using the emissivity in Eq.\,(15) of Persic \& Rephaeli (2019a), where the 
total (thermal) gas proton density is $n_g=n_{\rm HI}+2\,n_{\rm H_2}$ with the {average} neutral-hydrogen density $n_{\rm HI} 
= 0.54$ cm$^{-3}$, inferred from $M_{\rm HI} = 4.23 \, 10^8 M_\sun$ (this value of $M_{\rm HI}$ was deduced from direct 
determination of HI column density from the 21\,cm emission line; Stanimirovi\'c et al. 1999), and the molecular-hydrogen 
density $n_{\rm H_2} = 0.02$ cm$^{-3}$, inferred from $M_{\rm H_2} = 3.2 \, 10^7 M_\sun$ (which, in turn, is derived from 
modeling {\it Spitzer}, Cosmic Background Explorer (COBE), InfraRed Astronomical Satellite (IRAS), and Infrared Space Observatory 
(ISO) far-infrared data; Leroy et al. 2007). The CRp spectrum derived from our fit to the {\it Fermi}/LAT data is $N_{p0} = 
2.4 \, 10^{-10}$ cm$^{-3}$, $q_p = 2.40$, with $E_p^{max} = 30$ GeV. The model spectrum fully reproduces the 0.2--50 GeV {\it 
Fermi}/LAT spectrum (Fig.\ref{fig:gamma}). With these values, the estimated CRp energy density $u_p = \int_{m_p}^{E_p^{max}} 
E_p \,N_p(E_p) \, dE_p$ is $\sim 0.5$ eV cm$^{-3}$.

The closely related $\pi^\pm$-decay secondary CRe spectrum, $N_{se}(\gamma)$, has no free parameters once the CRp spectrum is 
determined
\footnote{
Denoting the total cross-section for inelastic $pp$ collisions $\sigma_{pp}(E_p)$ (see Eq.[79] of Kelner et al. 2006), the 
secondary CRe injection spectrum is $Q_{se}(\gamma) = (8/3) \,m_e \, (c / k_{\pi^\pm}) \, n_g\, N_{p0}\, [m_p + (4 m_e / 
\kappa_{\pi^{\pm}})\, \gamma]^{-q_p}\, \sigma_{pp}(m_p + E_{\pi^\pm}/k_{\pi^\pm})$ for $\gamma_{thr} < \gamma 
< \gamma_{se}^{max}$, with $\gamma_{se}^{thr} = m_{\pi^\pm}/(4m_e) = 68.5$ and $\gamma_{se}^{max}$$=$$m_{\pi^\pm}^{max}/(4m_e) 
\simeq (E_p^{max} - 3/2 m_p)/(4 m_e)$. The corresponding steady-state distribution is $N_{se}(\gamma) = 1/b(\gamma) \cdot 
\int_\gamma^{\gamma_{se}^{max}} Q_{se}(\gamma) \, d\gamma$, where $b(\gamma)$ is the radiative energy loss term. In a 
magnetized medium consisting of ionized, neutral and molecular gas, it is $b(\gamma) = \sum_{j=0}^2 b_j(\gamma)$ where 
$b_j(\gamma)$ are loss terms appropriate to each gas phase (see Eqs.[4]-[6] of Rephaeli \& Persic 2015). $N_{se}(\gamma)$ 
is analytically approximated by $N_{se}^{\rm fit}(\gamma) = N_{se,0} \gamma^{-q_1} (1+\gamma / \gamma_{b1})^{q_1-q_2} {\rm 
exp}[{-(\gamma/\gamma_{b2})^\eta}]$, where $q_1$, $q_2$ and $\gamma_{b1}$, $\gamma_{b2}$ are the low-/high-energy spectral 
indices and breaks; $\eta$ gauges the steepness of the high-end cutoff. 
}.
We use $N_{se}^{\rm fit}(\gamma)$, with parameter values reported in Table 4, to compute the corresponding leptonic yields. 
However, the uncertainty in the total (HI and H$_2$) gas density clearly affects the energy loss rate $b(\gamma)$ 
and the resulting spectral normalization and shape of $N_{se}(\gamma)$, hence also of $N_e(\gamma)$, and their yields. To 
compute the synchrotron emission we assume $B=3.5 \mu$G, based on estimates of the ordered (1.7\,$\mu$G) and random (3\,$\mu$G) 
fields in the SMC (Mao et al. 2008). 

Once the secondary-CRe synchrotron yield has been computed, its low-frequency residuals from the data are modeled using a CRe 
spectrum, $N_e(\gamma)$, with $N_{e0} = 3.7 \, 10^{-8}$ cm$^{-3}$, $q_e = 2.23$, $\gamma_{max}=10^4$. The latter is the 1PL 
approximation, for $100 \mincir \gamma \mincir \gamma_{max}$, of the actual steady-state primary-CRe spectrum. {\rm Neglecting 
diffusion and advection losses,} this primary spectrum is $N_{pe}(\gamma) = k_i \gamma^{-(q_i-1)} /[b(\gamma) \,(q_i -1)]$ 
cm$^{-3}$ (units of $\gamma$)$^{-1}$ where $\dot{N}_i(\gamma) = k_i\, \gamma^{-q_i}$ is the CRe spectral injection rate. We find 
that, over the mentioned electron energy range, $q_i=2.28$ provides a decent match between the shapes of the curved spectrum and 
the 1PL spectrum -- this value is typical for $\gamma$-ray emission from Galactic CR accelerators (e.g. SNRs, the Crab nebula). The 
relative normalization between the two spectra can be based on the measured flux density at some radio frequency or on imposing 
the same CRe energy density on the two spectra over the electron energy range of interest (see discussion in Rephaeli \& Persic 
2015). In our case here the normalization of $N_{pe}(\gamma)$, and hence $k_i$, can be found by using both $N_{pe}(\gamma)$ and 
$N_{se}(\gamma)$ to compute the total (primary plus secondary) synchrotron yield, which in turn is fitted to the radio synchrotron 
data: in doing this, the acceptable match (in the relevant energy range) between the curved and PL spectra allows us to use $N_e
(\gamma)$ without substantial loss of accuracy. Given this procedure, the uncertainties in 
the gas density ultimately affect $N_e(\gamma)$ as well.

The spectra of the synchrotron-emitting CRe are shown in Fig.\,\ref{fig:MC_CRe_spectra}. The resulting primary electron component 
is quite dominant, with $\sim$70\% of the total synchrotron and Compton yields (Fig.\,\ref{fig:SMC_radio}), but the normalization, 
$N_{e0}$, is not strongly constrained due to its dependence on the assumed magnetic field. If measurements of NT--X-ray and $\sim
$1-30 MeV $\gamma$-ray fluxes -- corresponding to, respectively, the CMB and CIB peaks -- become available, most of the uncertainty 
could be removed (e.g. Persic \& Rephaeli 2019a). In addition to this modeling uncertainty, the considerable coupling between the 
CRe spectral parameters implies an appreciable range of the deduced values of $N_{e0}$ and $q_e$: This range can be roughly bracketed 
by a flatter slope ($q_e=2.2$) and lower normalization ($N_{e0} = 2.9 \, 10^{-8}$ cm$^{-3}$) or, alternatively, a steeper slope 
($q_e=2.3$) and higher normalization ($N_{e0} = 6.7 \, 10^{-8}$ cm$^{-3}$) -- both, with the same $\gamma_{max}$ as the reference 
model (Table 4). 
 
At high frequencies a relatively flat, $\propto \nu^{-0.1}$, component represents diffuse thermal-ff emission (Spitzer 1978). The 
latter flux may be gauged to the H$\alpha$ flux if both emissions come from the same emitting volume (HII regions), because in this 
case the relevant warm-plasma parameters (temperature, density, filling factor) are the same. So the measured (optical) H$\alpha$ 
flux may be used to predict the (radio) ff emission. We model the thermal-ff emission combining Eqs.(3),(4a) of Condon (1992) and 
using $T_e = 1.1 \, 10^4$ $^{\circ}K$ (Toribio San Cipriano et al. 2017) and $F(H_\alpha) = 1.6 \, 10^{-8}$ erg cm$^{-2}$ 
s$^{-1}$ (Kennicutt et al. 1995). The (primary) synchrotron and thermal-ff normalizations are spectrally quite apart so they do not 
significantly affect each other. The radio model is shown alongside data in Fig.\,\ref{fig:SMC_radio}-right.

Although subdominant, secondary CRe contribute appreciably to the total CRe population of the SMC. Whereas the 
primary spectrum is approximately PL, the secondary spectrum is clearly curved (see Fig.\,\ref{fig:MC_CRe_spectra}). 
The resulting curvature of the total CRe spectrum is reflected in the total synchrotron spectrum (computed using Eq.\,(9) 
of Persic \& Rephaeli (2019a) for $N_e(\gamma)$ and its straightforward generalization for $N_{se}(\gamma)$). Therefore 
the total radio spectrum, which consists of synchrotron and thermal-ff emission at low and high frequencies, is not a 
smooth 2PL but shows some extra structure. A 3rd-order polynomial (in log units) outperforms For et al.'s (2018) 
''preferred'' 1PL model ($\Delta$BIC$>0$; Table 3). This polynomial (4 free parameters; Fig.\,\ref{fig:SMC_radio}-left) 
is matched by the physical model described above, i.e. a low-frequency component representing synchrotron emission (3 
free parameters) plus a high-frequency $\propto \nu^{-0.1}$ PL representing thermal-ff emission (1 free parameter); this 
model is shown in Fig.\,\ref{fig:SMC_radio}-right.


\begin{table*}
\caption[] {Summary of fits to the SMC radio spectrum.}
\centering 
\begin{tabular}{ l  l  l  l  l  l  l  l  l  l  l  l}
\hline
\hline
\noalign{\smallskip}
            &  ${\chi^2}^\bullet$  &  $\chi^2_{\rm red}$  & BIC$^\ddagger$ &  $\Delta$BIC  &  dof  & log$(S_0)$ &  $\alpha$  & $c_0$  &  $c_1$   & $c_2$ & $c_3$ \\
\noalign{\smallskip}
\hline
\noalign{\smallskip}
PL$^\star$  &    32.3              &           1.2        & 298.0          &               &   27  &   2.284    &  $-0.82$   &        &          &       &       \\
P$_3^+$     &    22.3              &           0.9        & 294.7          &     3.3       &   25  &            &            &  1.763 & $-$0.968 & 0.09  & 0.315 \\

\noalign{\smallskip}
\hline\end{tabular}
\smallskip

$^\star$ Power law: ${\rm log}(S_\nu) = {\rm log}(S_0) + \alpha\, {\rm log}(\nu/0.2\,{\rm GHz})$ from Eq.(4) and Table 4 of For et al. (2018).
\smallskip

\noindent
$^+$ 3rd-order polynomial: ${\rm log}(S_\nu) = c_0 + c_1\,x + c_2\,x^2 + c_3\,x^3$, with $x = {\rm log}(\nu/{\rm GHz})$.
\smallskip

\noindent
$^\bullet$ $\chi^2$ calculations use actual fluxes (Table 3 of For et al. 2018), not their logarithms. 
\smallskip

\noindent
$^\ddagger$ See definitions and discussion in Sect. 4.2.1 of For et al. 2018. 

\smallskip
\end{table*}

With the CRe spectra essentially determined, we can calculate the Compton and NT-bremsstrahlung yields from CRe scattering 
off, respectively, CMB/EBL/GFL photons and thermal-plasma nuclei using, respectively, Eqs. (2.42) and (3.1) of Blumenthal \& 
Gould (1970). The photon fields are described in section 3; as to the densities of plasma nuclei we assume $n_i = 0.033$ 
cm$^{-3}$ (Mao et al. 2008) and metallicity $Z^2=1.2$ (Sasaki et al. 2002). Although the shape of the $\gamma$-ray spectrum 
is distinctly pionic, NT-bremsstrahlung peaks near the pionic peak ($\nu \sim 10^{23}$ Hz) where it contributes $\sim$3\% 
of the measured flux; Compton/starlight emission, that peaks at $\nu \sim 10^{22}$ Hz, contributes $\mincir$1\% of the 
flux at the lowest LAT data point. The predicted Compton/CMB spectral flux, too, is shown in Fig.\,\ref{fig:SMC_SED}. 
We found no need to introduce a further, broken-PL component in the $\gamma$-ray band representing pulsars, in agreement 
with Lopez et al. (2018).

The resulting SED model, overlaid on data, is plotted in Fig.\ref{fig:SMC_SED}. 

An upper limit on the mean magnetic field, $B_{max}$, can be estimated by assuming that the measured radio emission is fully 
produced by secondary CRe. By matching the measured and predicted emission levels we infer a limiting value $B_{lim} \simeq 7 
\, \mu$G. This upper limit, which is twice the observationally deduced value, is unrealistically high because, most notably 
at low frequencies, the {\it shapes} of the measured and predicted synchrotron spectra do not match well, and because primary 
CRe obviously contribute to the measured radio flux. A lower limit, $B_{min}$, can estimated from the fact that if $B$ decreases 
the secondary synchrotron yield decreases too, so to keep the synchrotron yield unchanged, the primary yield must increase -- by 
increasing the spectral normalization ($N_{e0}$) and possibly also varying the spectral shape ($q_e$, $\gamma_{max}$) of the 
primary CRe spectrum. At low energies the primary spectrum is steeper than the secondary, so at some point the resulting 
synchrotron spectrum becomes progressively and systematically steeper than the $<$230\,GHz radio data; this yields $B_{min} 
\simeq 2\,\mu$G. Based on our SED analysis, the value $B=3.5 \mu$G suggested by Mao et al. (2008) lies comfortably between 
the estimated bounds.

%
\begin{figure*}
\vspace{6.2cm}
\includegraphics{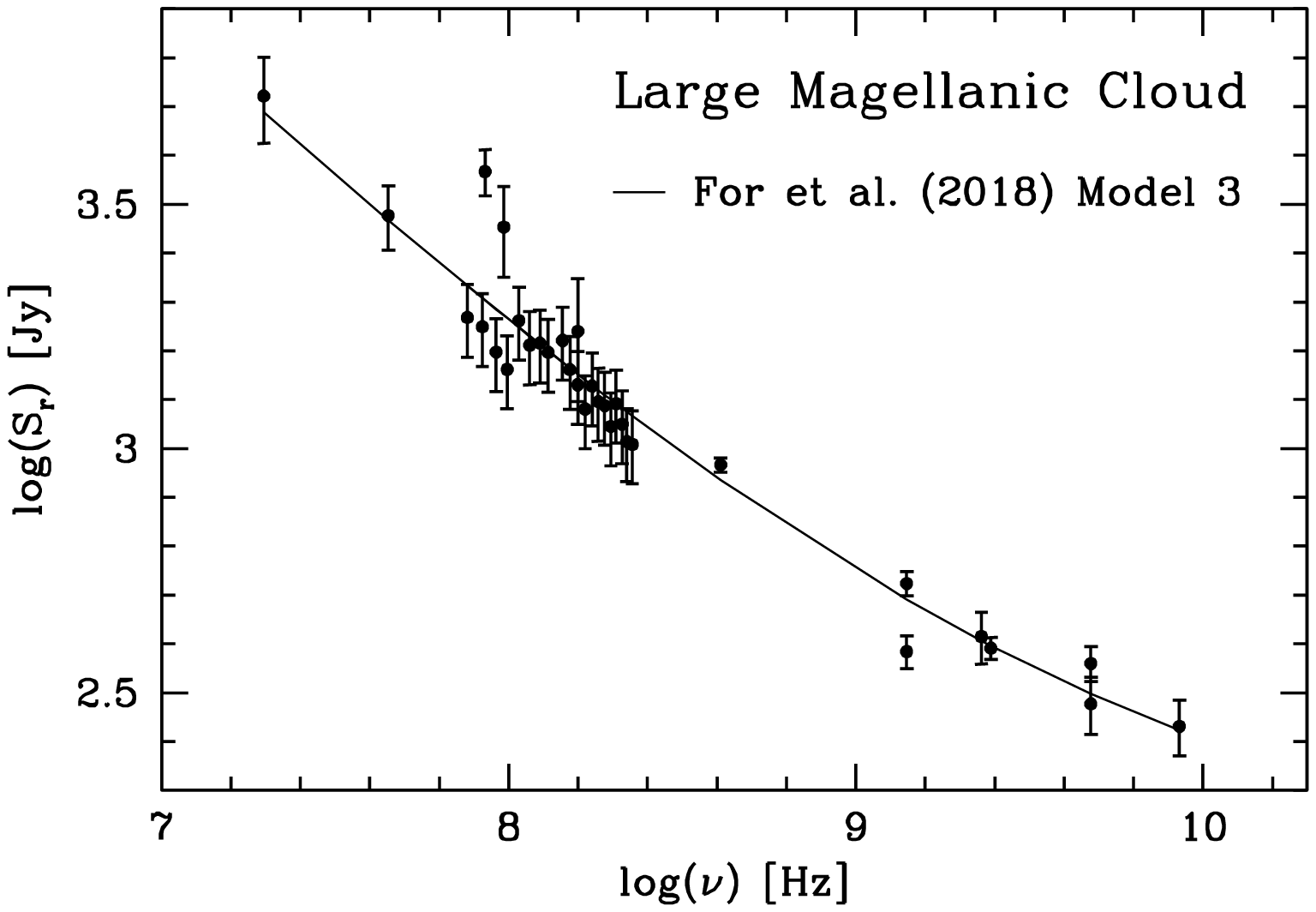}
\includegraphics{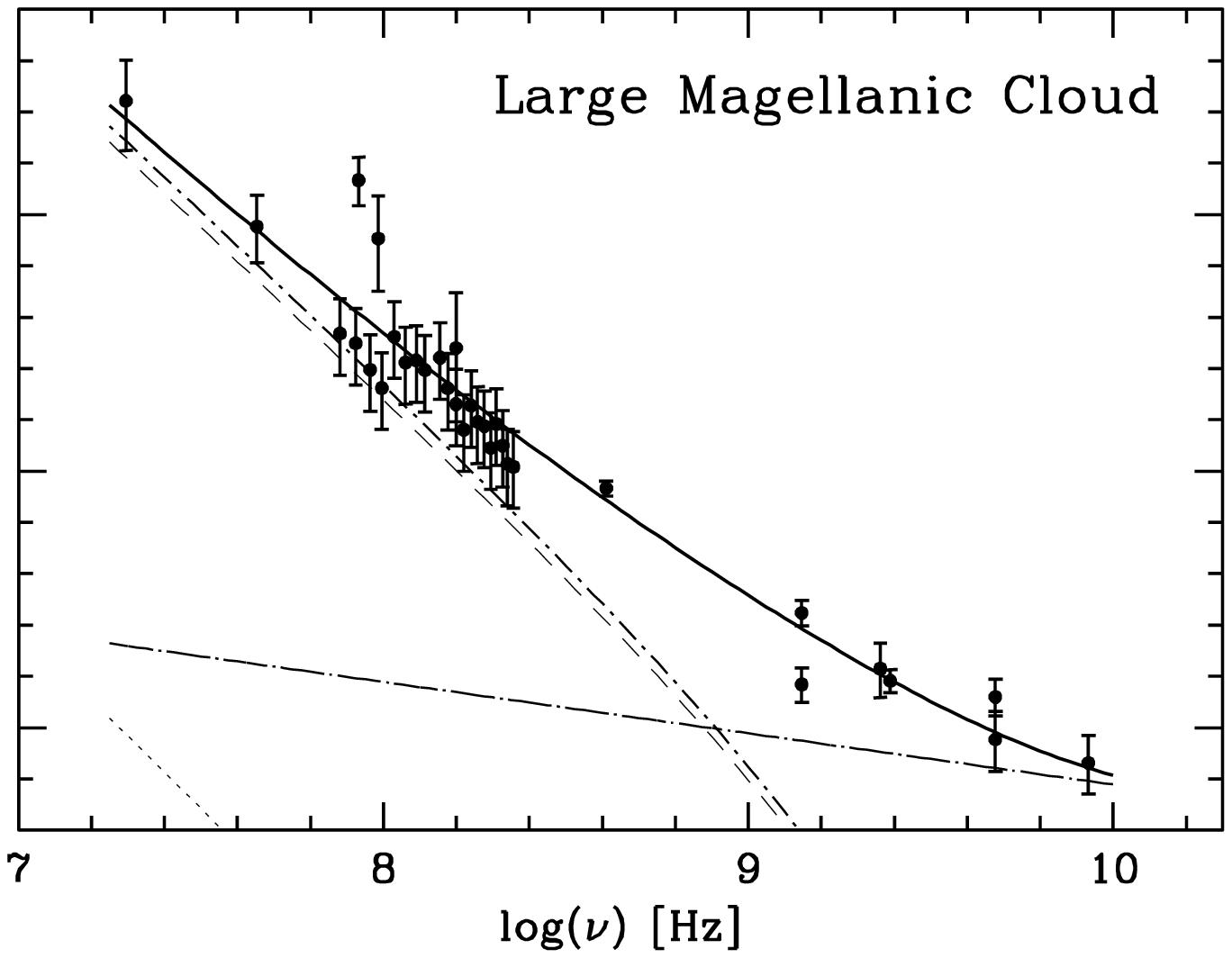}
\caption{ 
LMC radio spectrum: data points (dots) are from Table 2 of For et al. (2018). 
{\it Left:} the 2PL model\,3 of For et al. (2018) is shown as a solid curve.
{\it Right:} the two-component model (solid curve) includes synchrotron radiation (dotted/short-dashed curve; 
dashed and dotted curves denote, respectively, primary and secondary contributions - the latter corresponding 
to T+17 $\gamma$-ray data) and thermal-ff emission (dotted/long-dashed curve). 
}
\label{fig:LMC_radio}
\end{figure*}
%

%
\begin{figure}
\vspace{10.0cm}
\includegraphics{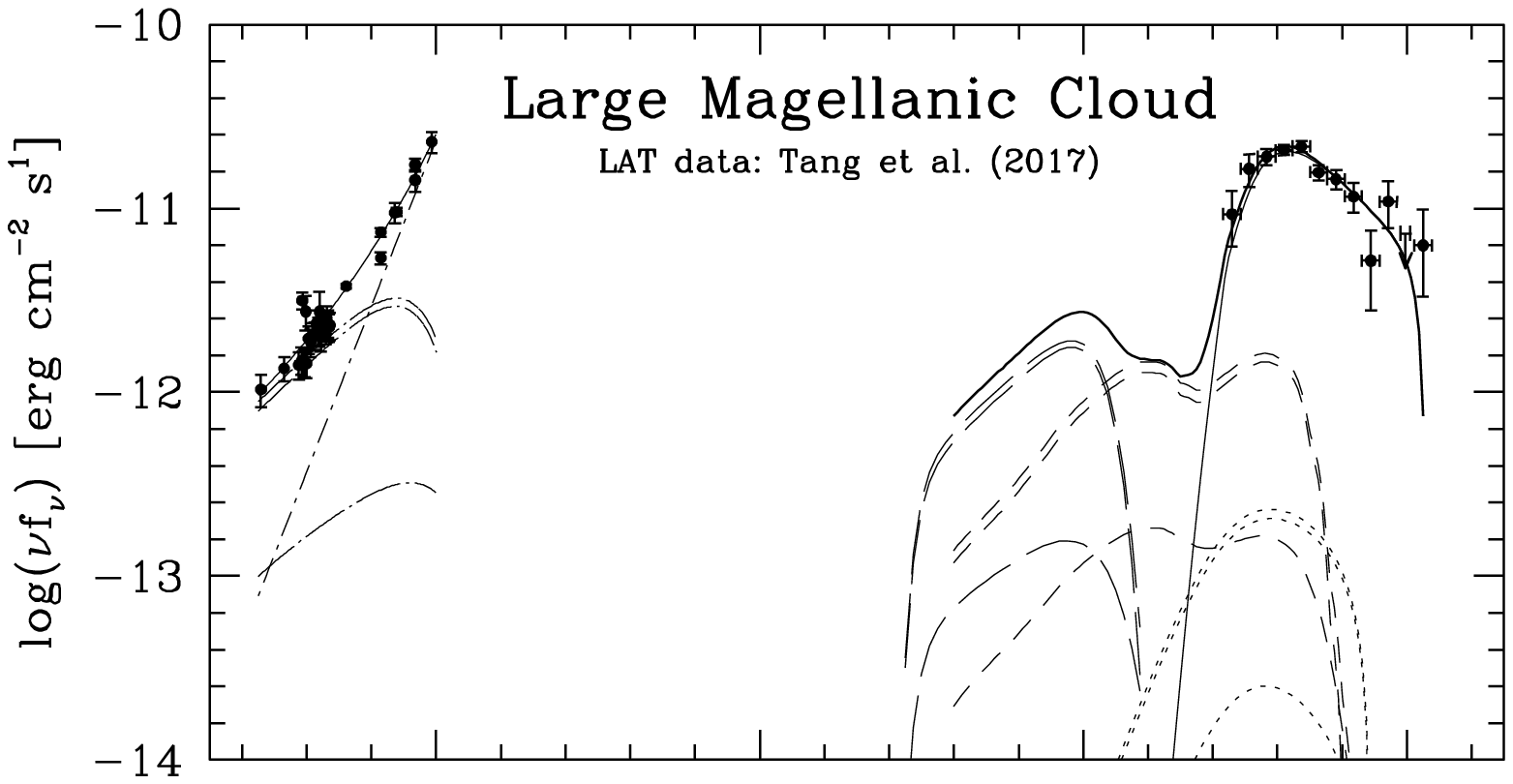}
\includegraphics{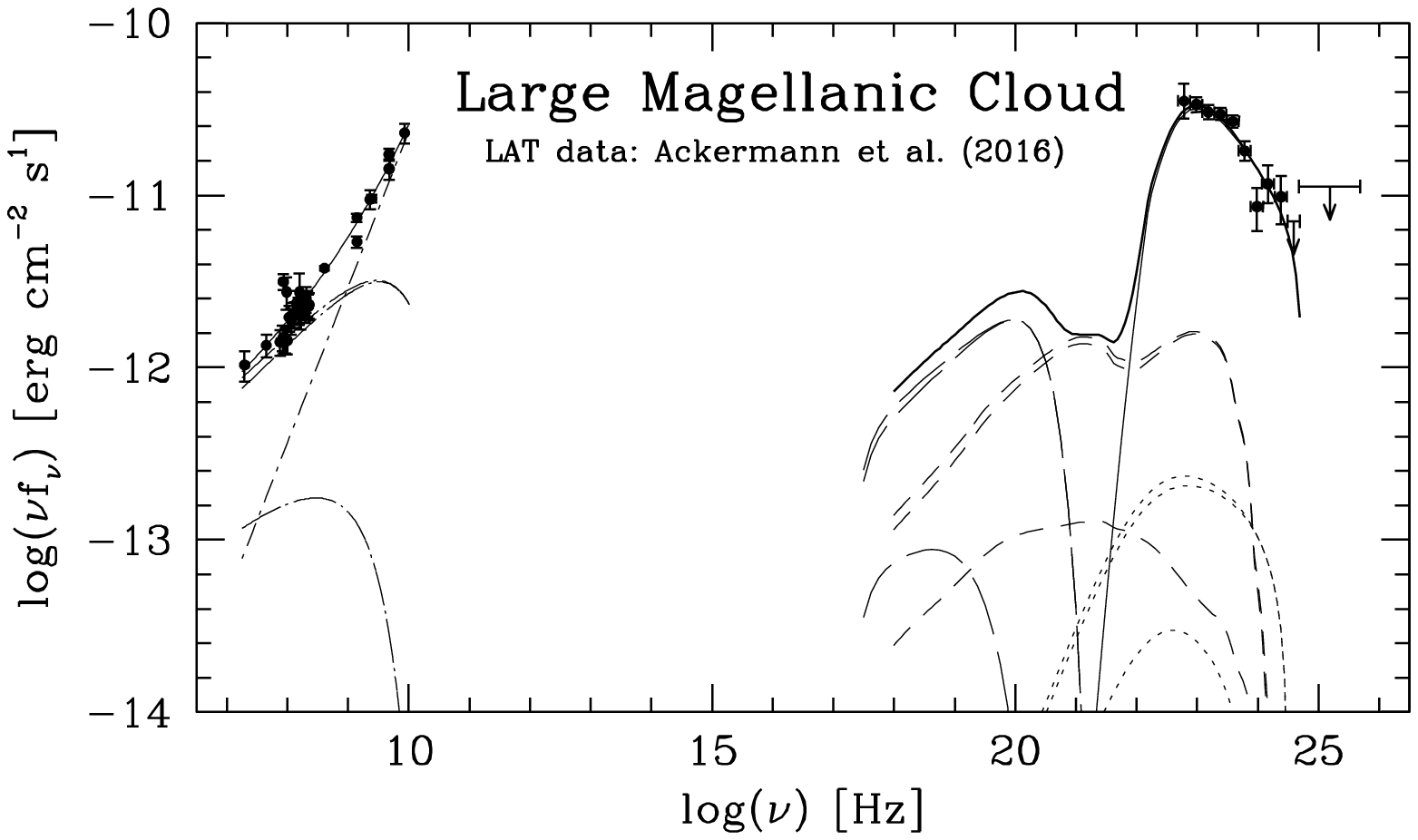}
\caption{ 
The broad-band (radio/$\gamma$, 19.7\,MHz -- 100\,GeV) SED of the LMC disk. 
The $\gamma$-ray data are from T+17 ({\it top}; reference set) and 
A+16 ({\it bottom}; auxiliary set). Symbols are as in Fig.\,\ref{fig:SMC_SED}
}
\label{fig:LMC_SED}
\end{figure}
%

\subsection{LMC} 

Our fitting approach differs from that adopted for the SMC. The fact the radio spectrum is 
best fitted by a 2PL model interpreted as a combination of (individually 1PL) synchrotron and thermal-ff components (For 
et al. 2018), indicates that synchrotron emission is generated by a 1PL CRe spectrum. This is unlike the case of the SMC 
for which the interplay of primary and secondary CRe with different spectra results in a more structured, non-PL form. 
With the primary CRe spectrum determined from fitting the radio data, the Compton $\gamma$-ray yield is calculated. To 
this a modeled pionic component is added and fit to the $\gamma$-ray data. Doing so enables determination of the the 
secondary CRe spectrum and the resulting yields, which turn out to be very minor in comparison with those of primary CRe; 
thus, no iteration of the fitting process is required.

The radio flux is dominated by synchrotron emission with $\alpha = 0.66$ at the lowest measured frequencies, and by 
thermal-ff emission with $\alpha = -0.1$ at the highest frequencies (For et al. 2018). Adopting these spectral indices 
for the two components, we fit the full radio database with $T_e = 1.3 \,10^4$ $^{\circ}K$ and $F(H\alpha) = 2\, 10^{-7}$ 
erg cm$^{-2}$ s$^{-1}$ (Toribio San Cipriano et al. 2017; Kennicutt et al. 1995) for the thermal-ff component, and $B = 
4.3 \mu$G (Gaensler et al. 2005)
\footnote{
This field value, from a Faraday rotation study of the LMC, is consistent with Mao et al.'s (2012) 
estimates, $B_{\rm eq} \sim 2\mu$G from equipartition with CRp assuming $q_p=2$ and $B < 7\mu$G from 
using a (deduced) lower limit on $N_{e0}$ in fitting the 1.4 GHz radio flux.
},
$N_{e0} \simeq 3.5\, 10^{-7}$ cm$^{-3}$ and $\gamma_{\rm max} \simeq 2.4\, 10^4$ for the synchrotron. (The cutoff is 
needed for consistency with the measured flux at $\nu \magcir 3\, 10^{23}$ Hz.) We should mention at this point that 
subsequent results (see below) suggest that the secondary CRe are only $\sim$10\% of the total. So the (quasi-)1PL 
primaries dominate the CRe population (Fig.\,\ref{fig:MC_CRe_spectra}) and the emerging synchrotron emission. Therefore, 
the combined synchrotron and thermal-ff radio spectrum matches the smooth 2PL profile of For et al.'s (2018) fit. The 
radio model is shown in Fig.\,\ref{fig:LMC_radio}. 

Next, we calculate the X/$\gamma$-ray Compton and NT-bremsstrahlung yields from {\rm (primary)} CRe scattering off 
CMB/EBL/GFL photons and thermal-plasma nuclei: {\rm to this aim we assume} the photon fields described in section 3 
{\rm and, respectively, a plasma characterized by an average density} $n_i = 0.012$ cm$^{-3}$ (Sasaki et al. 2002; 
Cox et al. 2006) and metallicity $Z^2=1.2$ (Sasaki et al. 2002). {\rm Both emissions fall short of reproducing the 
$\gamma$-ray data.}

The pionic emission is computed with $n_{\rm HI}=1$ cm$^{-3}$ (from $M_{\rm HI}=3.8\, 10^8 M_\sun$: Staveley-Smith et 
al. 2003) and $n_{\rm H_2} = 6.6\, 10^{-2}$ cm$^{-3}$ (from $M_{\rm H_2}=5\, 10^7 M_\sun$, measured from CO emission: Fukui 
et al. 2008). From our fitting, the matching CRp spectrum has $q_p = 2.38$ and $E_p^{max} = 80$ GeV for the T+17 data 
(Fig.\,\ref{fig:LMC_SED}-{\it top}), and $q_p = 2.6$ and $E_p^{max} = 25$ GeV for the A+16 data (Fig.\,\ref{fig:LMC_SED}-{\it bottom}). 
The corresponding CRp energy densities are, respectively, 1 and 1.5 eV cm$^{-3}$ -- the difference being largely due to 
the somewhat higher flux level in the A+16 data. In spite of the observational uncertainties in the two datasets, we see 
that for either the pionic yield largely dominates the $\gamma$-ray emission, so the pionic nature of the measured 
$\gamma$-ray spectrum appears well established. In addition to the dominant pionic component a subdominant, 7\% (5\%), 
leptonic contributions peaks at 500 (400) MeV, similarly to the pionic one: Compton/COB and NT bremsstrahlung emissions 
contribute, respectively, 6.5\% (4\%) and 0.5\% (1\%) to the peak of the observed emission in the T+16 (A+16) database. 

The secondary CRe spectrum is derived and fitted analytically as described in section 4.1: the fitting parameters are 
reported in Table 4 for both sets of LAT data. As mentioned, in both cases the radiative contribution of the secondary 
CRe is minor: this means that the corresponding primary PL spectra are essentially the same 
\footnote{
For the adopted nominal densities of the gas and of the magnetic and radiation fields, the inferred primary CRe 
injection index in the LMC is $q_i=2.22$ for both the T+17 and A+16 fits. 
}
and (for each data set) no fitting iteration is required. The primary and secondary CRe spectra are shown in 
Fig.\,\ref{fig:MC_CRe_spectra}. 

Primary CRe largely dominate the CRe population in the LMC, even more so than in the SMC. However, their density 
is effectively unconstrained because it depends on assuming a magnetic field value (even though deduced from, as 
in our case, Faraday-rotation measurements) rather than using a value derived by modeling the NT--X-ray emission 
as Compton/CMB (e.g., Persic \& Rephaeli 2019a,b) or the hard--X-ray/MeV emission (when it becomes  available) as 
Compton/IR. Also, appreciable modeling uncertainty stems from the inter-dependence among primary-CRe parameters 
which results in a range of acceptable values of the spectral index and normalization, $q_e=2.26$ with $N_{e0} = 
2.22\, 10^{-7}$ cm$^{-3}$ and $\gamma_{max} = 2.0\, 10^4$, as well as $q_e=2.40$ with $N_{e0} = 6.21\, 10^{-7}$ 
cm$^{-3}$ and $\gamma_{max} = 2.7\, 10^4$.
\footnote{
Values referring to the T+17 LAT data. A similar covariance, with only 
slightly different values, is found when the A+16 LAT data are used.
}
%


\begin{table*}
\caption[] {SED model parameters.}
\centering 
\begin{tabular}{ l  l  l  l  l  l  l  l  l  l  l  l  l  l  l}
\hline
\hline
\noalign{\smallskip}
      &   $N_{e0}$          &$q_e$ & $\gamma_{max}$&  $u_p$ &$q_p$&$E_p^{max}$     &  $N_{se0}$              &$q_1$&$q_2$&$\gamma_{b1}$&$\gamma_{b2}$  &$\eta$& F(H$\alpha$) & $T_e$    \\
    &{\tiny cm$^{-3}$}      &      & {\tiny $10^4$}&{\tiny eV/cm$^3$}& &{\tiny GeV}&{\tiny $10^{-11}$cm$^{-3}$}&   &     &{\tiny $10^2$}&{\tiny $10^4$}& &{\tiny erg/(cm$^2$s)}&{\tiny $10^{4\,\circ}K$}\\
\noalign{\smallskip}
\hline
\noalign{\smallskip}
SMC         &$3.70~ 10^{-8}$& 2.23 &  $1.0$   &   0.45   & 2.40&    30     &$0.14$ & 0.10& 2.60&     1.5    & $0.95$ & 2.65 & $1.6~10^{-8}$&$1.1$\\
LMC$^+$ [1] &$3.50~ 10^{-7}$& 2.32 &  $2.4$   &   1.00   & 2.38&    80     &$0.50$ & 0.05& 2.67&     1.2    & $2.75$ & 3.20 & $2.0~10^{-7}$&$1.3$\\
~~~~~~~~~~~ [2] &$3.30~10^{-7}$& 2.32& $2.6$  &   1.55   & 2.60&    25     &$1.55$ & 0.13& 2.75&     1.0    & $0.85$ & 3.20 & $2.0~10^{-7}$&$1.3$\\

\noalign{\smallskip}
\hline\end{tabular}
\smallskip

\noindent
$^+$ {\it Fermi}/LAT $\gamma$-ray data from T+17 [1] and A+16 [2]. 
\smallskip

\end{table*}

\section{Neutrino emission} 

\begin{figure*}
\vspace{6.2cm}
\includegraphics{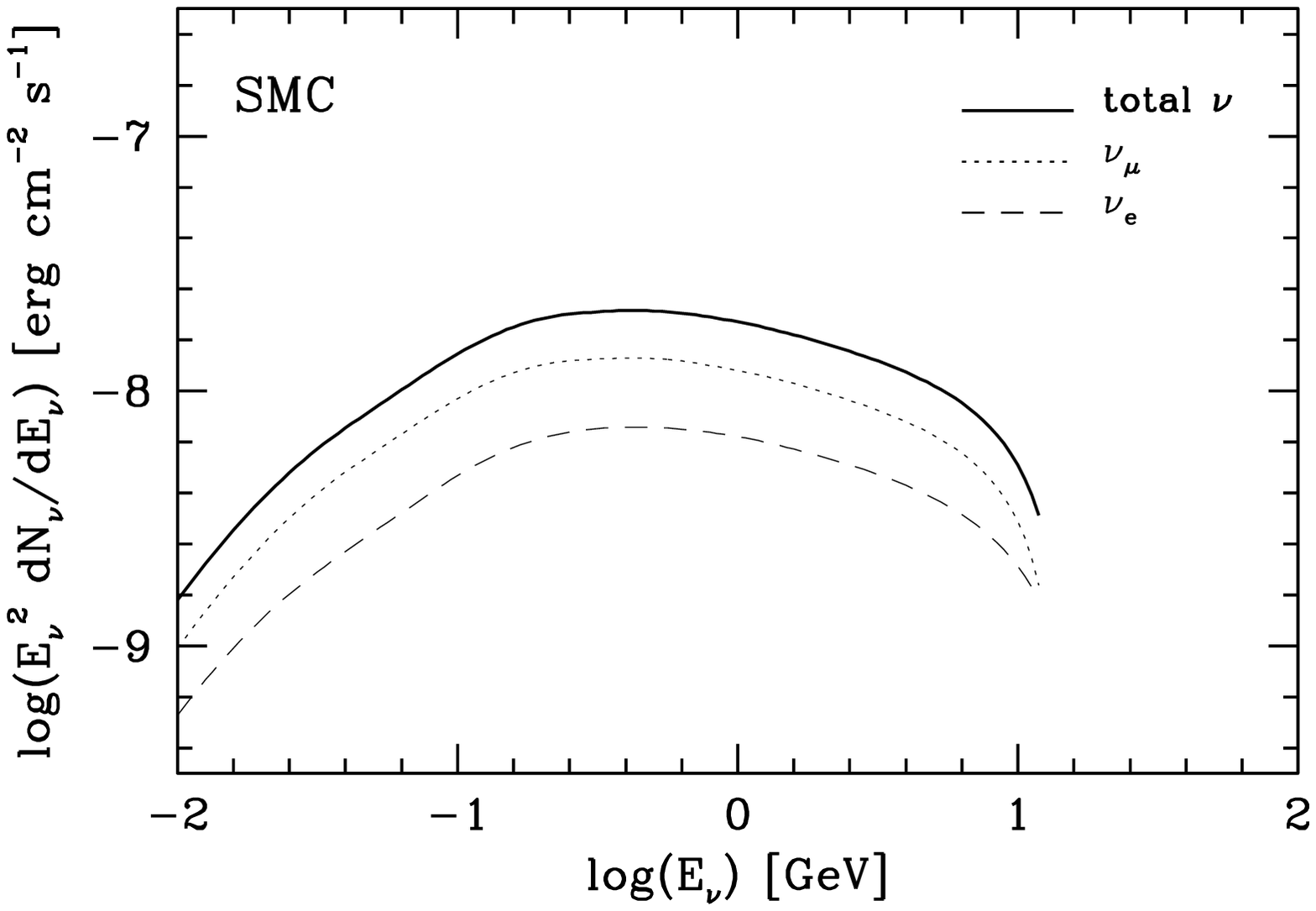}
\includegraphics{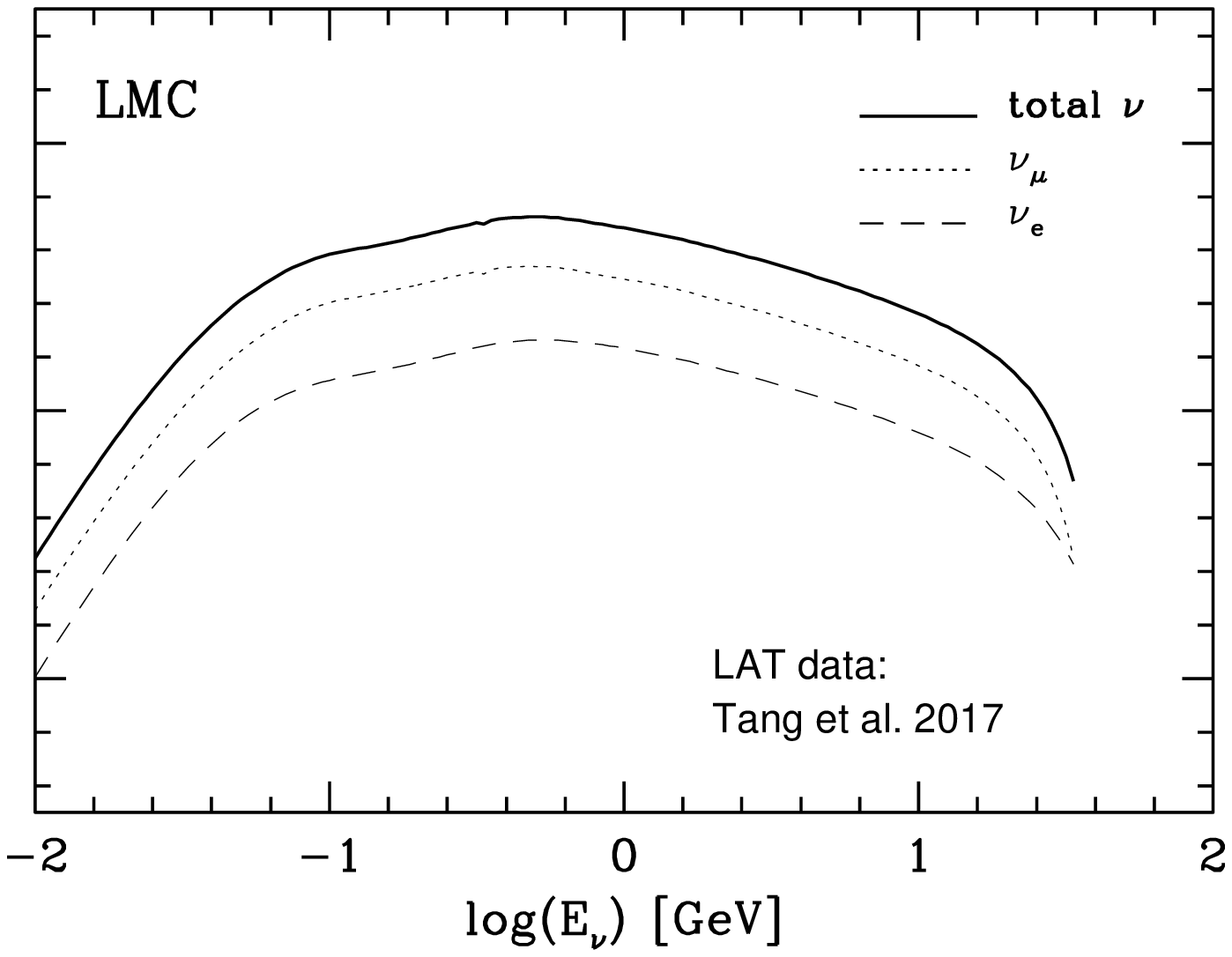}
\caption{ 
Neutrino spectral emissions in the Magellanic Clouds. The LMC panel refers to the T+17 {\it Fermi}/LAT dataset.
}
\label{fig:MC_neutrinos}
\end{figure*}

With an apparent dominant $\pi^0$-decay origin of the $\gamma$\,rays produced in the interstellar medium of both MCs it is clear that 
$\pi^\pm$-decay neutrinos are also produced. Our calculations (following Kelner et al. 2006) of the predicted muon- and electron-neutrino 
spectra of both galaxies indicate (Fig.\,\ref{fig:MC_neutrinos}) that the broadly-peaked ($\sim$0.1--10 GeV) neutrino flux is too low 
for detection by current and upcoming $\nu$ projects. This conclusion is based on the estimated observation time needed to detect the 
LMC with an experiment with detection sensitivity comparable to the Antarctica-based IceCube+DeepCore Observatory, the most sensitive 
current/planned $\nu$-detector at neutrino energies $10 < E_\nu/{\rm GeV} <100$ (e.g. Bartos et al. 2013). The latter's effective area 
(Abbasi et al. 2012) is $A_{\rm eff}(E_\nu) = 40 \left( \frac {E_\nu}{100\,{\rm GeV}}\right)^2$ cm$^2$ (for $\nu_{\mu}$, twice smaller 
for $\nu_e$) in this energy range (Bartos et al. 2013). Only in the narrow energy range 10-50\,GeV do the IceCube+DeepCore sensitivity 
and the LMC predicted diffuse spectral $\nu$-flux effectively overlap (cf. the T+17 LAT dataset). In this band the latter can 
be approximated as $\frac{dN_\nu} {dE_\nu} \sim 10^{-10} \left( \frac{E_\nu}{100\, {\rm GeV}} \right)^{-2.5} {\rm cm}^{-2} {\rm s}^{-1} 
{\rm GeV}^{-1}$. The corresponding number of detected neutrinos, $N_\nu = t_{\rm obs} \int_{10\,{\rm GeV}}^{50\,{\rm GeV}} \frac{dN_\nu}
{dE_\nu} A_{\rm eff}(E_\nu) dE_\nu$, is $N_\nu \sim 10 (t_{\rm obs}/{\rm yr})$. The detector background (for up-going events) is dominated 
by atmospheric neutrinos produced by cosmic rays in the northern hemisphere; its energy spectrum is approximately flat in the relevant 
energy range, at a level $\sim$$10^2 (d\Omega_{\rm LMC} / 1.6\, {\rm sr})$ Gev$^{-1}$ yr$^{-1}$ (Bartos et al. 2013 and references therein). 
So the net 10-50\,GeV background rate is $\sim$$4 \times 10^3 (\Delta\Omega_{\rm LMC} / 1.6\, {\rm sr})$ yr$^{-1} \sim 60$ yr$^{-1}$ 
assuming the LMC angular radius to be 5 deg). Based on this crude estimate, observation of diffuse GeV neutrinos from the LMC disk would 
imply $S/N < 0.2$.

%
%

\section{Conclusion} 

The SMC and the LMC, star-forming Galactic satellite galaxies, are among the brightest sources in the {\it Fermi}/LAT $\gamma$-ray 
sky. We self-consistently modeled the radio/$\gamma$ SED of both galaxies using the latest available radio and LAT data, using exact 
emissivity formulae for the relevant emission processes. Both SEDs were modeled with the radio data interpreted as a combination of 
NT electron synchrotron emission and thermal electron bremsstrahlung, and the $\gamma$-ray data as a combination of $\pi^0$-decay 
emission from CRp interacting with the ambient gas plus leptonic emission. The CRp spectra appear similar in the two galaxies with 
$q_p \sim 2.4$, and $u_p \sim 1$ eV cm$^{-3}$. Our spectral findings are qualitatively and quantitatively new for the SMC, and confirm 
and strengthen previous results for the LMC. In detail, we quantify Lopez et al.'s (2018) suggestion of a mostly pionic origin of the 
LAT-measured $\gamma$-ray emission of the SMC, only $\mincir$3\% of the peak flux being by NT bremsstrahlung; as to the LMC, our 
analysis suggests that 93\%, 6.5\%, and 0.5\% of the LMC peak $\gamma$-ray flux (at 0.5 GeV) are accounted for by pionic, 
Compton/(EBL+GFL), and NT-bremsstrahlung emission -- in broad agreement with Foreman et al. (2015). As we have noted, for the SMC disk 
we used Lopez et al.'s (2018) LAT dataset which is only mildly contaminated by an estimated $\sim$10\% contribution from unresolved 
pulsars, whereas for the LMC disk we used LAT datasets that are based on maps that do not include emission from local gas and CR 
inhomogeneities associated with actively star-forming regions (A+16; T+17). Thus, our pionic emission modeling results are unlikely 
to be appreciably affected by emission from individual sources. 

As stated in Section 4, our treatment is based on determining the particle steady-state spectral distributions in 
the disk by fits to the radio and $\gamma$-ray data, using previously deduced mean disk values of the gas density, 
magnetic and radiation fields. An assessment of the combined uncertainties in the predicted particle radiative (and 
neutrino) yields is based firstly on the precision level of the observationally determined values of $n_g$ and $B$. 
Whereas the deduced CRe normalization is (essentially) independent of $n_g$, its dependence on $B$ is significant, 
$N_{e0} \propto B^{-(\alpha +1)}$, which is roughly $B^{-1.7}$. The deduced CRp normalization is $N_{p0} \propto 1/n_g$. 
Given the substantial uncertainties in both $n_g$ and $B$, it is clear that these impact also the overall normalization 
of the particle spectra, which can be uncertain by a factor of 2-3. However, the constraining power of the fits to the 
measurements is mostly in the well known spectral shapes of the predicted synchrotron and $\pi^0$-decay processes. As 
explained in the previous section, the spectral fits are quite good, with a typical level of uncertainty of a few $10\%$. 
Even though substantial, the overall level of uncertainty is not large enough to question our qualitative conclusion on 
the pionic nature of the $\gamma$-ray emission. 

The SMC radio data have large error bars below 230 MHz where $q_e$ would be best determined, are unsampled (but for one point) between 
230 and 1400 MHz where the effect of the spectral truncation would be observed, and are only sparsely sampled above 1400 MHz. This 
leaves both $q_e$ and $\gamma_{max}$ poorly determined. The LMC radio spectral data permit determination only of $q_e$, as no truncation 
is apparent; we estimated $\gamma_{max}$ by weighing its lower limit implied by the radio spectrum and its upper limit implied by the 
$\gamma$ spectrum. Indeed, a larger value would shift the Compton/(EBL+FGL) blue peak to higher frequencies and to a higher flux level, 
deteriorating the fit to the $\gamma$ spectrum at frequencies log$(\,\nu)>24$. Our joint analysis of the broad-band SED of the MCs using 
exact emissivity formulae, and the leptonic radio and $\gamma$ emissions are coupled and rest on independent measures (independent of 
particle/field energy equipartition considerations) of the magnetic field, so the pionic emission is formally determined by modeling the 
residuals of the LAT data after the leptonic yield is accounted for. In general, uncertainties on the CRe spectrum affect the determination 
of the CRp spectrum because the latter is obtained by fitting the $\gamma$ data once the leptonic yields have been properly accounted for. 
However, we feel our CRp results here are reasonably safe because of the clear dominance of pionic emission in the $\gamma$ spectrum. 

For the SMC, the LAT data are sufficient to define the CRp spectral parameters, $q_p = 2.4$ and $E_p^{max} = 30$ GeV. For the 
LMC, however, some inconsistency between the T+17 and A+16 LAT datasets implies the derived CRp spectrum to be, respectively, 
flatter ($q_p = 2.4$) with no clearly discernible high-energy truncation ($E_p^{max} \mincir 80$\,GeV), or steeper ($q_p = 2.6$) 
with a clear truncation ($E_p^{max} = 25$\,GeV) and a $\sim$50\% higher normalization. (Both $q_p$ values are consistent within 
errors with Foreman et al.'s (2015) earlier estimate.) The reason for the discrepancy between the two data sets is not obvious, 
as the T+17 and A+16 data extraction regions largely overlap with each other and with the LMC disk, and both have been cleaned 
for identified (point-like or extended) sources. It should be pointed out that in the T+17 dataset the quality of the four 
highest-frequency points does not allow a definite estimate of $E_p^{max}$, whereas the four lowest-frequency points enable a 
clear view of the rising portion of the pionic hump. Ongoing {\it Fermi}/LAT measurements would likely improve the quality of 
the spectrum and clarify this issue. For both galaxies we obtained truncation energies systematically higher for CRp than for 
CRe -- probably, because energy losses are much less efficient for the former than for the latter in the relatively low-density 
MC interstellar gas.

The secondary CRe spectra are at best as well determined as those of the parent CRp. Our model secondary spectra (with no free 
parameters) are relatively well defined in the MCs. The primary CRe spectra, determined by fitting radio data after secondary CRe 
have been accounted for, are however less well constrained. First, the measured radio fluxes include emission from individual 
background and MC-disk point sources that contribute $\sim$20\% of the total emission and are spectrally similar to the overall 
emission (For et al. 2018), so there is some intrinsic uncertainty in the predicted extended emission's spectral shape and 
normalization ($N_{e0}$ is biased high). Secondly, because of the $q_e$-$N_{e0}$ degeneracy the uncertainty on $N_{e0}$ affects 
the determination of the spectral slope. Thus, primary CRe are rather poorly constrained in both MCs. In spite of this, our 
assumption of a (truncated) 1PL representing primary CRe is sound. The PL is a fair approximation, over the relevant CRe energy 
range, to the actual spectrum computed accounting for realistic energy losses in the MC disk environments. The approximation is 
particularly good for the LMC. 

A possible underestimate of the gas content in the MC due to the presence of "dark" (unseen) neutral gas, i.e. HI gas optically thick 
in the 21\,cm line and/or H$_2$ has with no associated CO emission, could increase the uncertainty on the total, atomic and molecular, 
gas density. The CO (J=1-0) transition, often used to detect the total H$_2$ mass, may be hard to detect in low-$Z$ galaxies such as 
the MC (Rolleston et al. 2002; Requena-Torres et al. 2016), so the relatively bright [C\,II]$\lambda$158\,$\mu$m line can be used 
instead: some $>$70\% may not have been traced by CO (1-0) in dwarf galaxies but is well traced by [C\,II]$\lambda$158\,$\mu$m (Madden 
et al. 2020). Another important aspect to investigate is the mixing of dust into the interstellar medium (ISM) and the spatial 
variations of their properties (gas clumpiness affects emission levels, see above). Comparing the distributions of the IR dust emission 
and several gas tracers (H$\alpha$, 21\,cm, CO emission), their different emission processes highlight the distribution of gas under 
different conditions. In the LMC, this analysis reveals that: 
{\it (i)} dust emission, sampled in the Multiband Imaging Photometer for {\it Spitzer} (MIPS) 70 and 160\,$\mu$m bands, is well mixed 
with the large-scale HI 21\,cm emission; 
{\it (ii)} H$\alpha$ from star-forming H\,II regions is confined to the massive star formation region where also warmer ($\sim$120\,K) 
24\,$\mu$m emission is found; and 
{\it (iii)} the {\it Spitzer} InfraRed Array Camera (IRAC) 8\,$\mu$m band, that traces polycyclic aromatic hydrocarbons, correlates well 
with the HI gas but is absent from the massive-star--forming H\,II regions. All in all, the dust emission revealed by the combined IRAC 
8\,$\mu$m and MIPS 24, 70 and 160\,$\mu$m bands traces all three phases of the ISM gas (Meixner et a. 2006). Underestimating the gas 
content would affect the particle 
spectra in the following ways: 
{\it (i)} overestimate the CRp spectral normalization; 
{\it (ii)} underestimate the Coulomb and bremsstrahlung energy losses, $b_0(\gamma)$ and $b_1(\gamma)$ (Rephaeli 
\& Persic 2015), which would cause the computed steady-state secondary-CRe spectrum to be, respectively, higher 
and steeper; 
{\it (iii)} bias the derived primary-CRe spectrum lower and flatter.

Uncertainties in the 3D structure of the MCs (mainly for the SMC, Abdo et al. 2010b) directly affect only the CRe and CRp 
spectral normalizations, to match the spectral flux when the emitting volume changes -- this holds for the relevant leptonic 
emissivities, which depend linearly on the CRe normalization. Thus, model SEDs are nearly unaffected by variations of the MC 
structure parameters. Another source of uncertainty in $N_{e0}$ stems from assuming a magnetic field (for both galaxies), in 
our case field values derived by Faraday rotation studies of extragalactic polarized sources seen through the MCs. These 
field strengths are measured along the line of sight to the background sources; thus, they are line-averaged fields, whereas 
the field in the synchrotron emissivity expression is a volume average. The two values may or may not be the same: the potential 
mismatch introduces another systematic uncertainty in the determination of $N_{e0}$ from fitting the radio spectrum. For 
example, in principle if $B$ is biased high, $N_{e0}$ and all leptonic yields are biased low, so the resulting $N_{p0}$ is 
biased high. 

The main breakthrough needed in MC SED studies is the measurement of diffuse NT X-ray emission from their disks. 
As exemplified in our studies of radio-lobe SEDs (Persic \& Rephaeli 2019a,b), the NT 1\,keV flux, interpreted as 
Compton/CMB emission, sets the normalization of the CRe spectrum: in the MC case, given that the excellent pionic 
fit to the $\gamma$-ray emission firmly defines the secondary spectrum, the 1\,keV flux would effectively measure 
the primary normalization. Modeling the radio spectrum as synchrotron radiation would then provide the primary 
spectral shape -- and, importantly, the magnetic field. A spectral determination of $B$ would bypass the need to 
assume particles/field equipartition, and would return a volume-averaged value better representing the mean field 
than the line averages yielded by Faraday-rotation--based measurements. Deeper X-ray observations of both MCs, with 
better spatial/spectral resolution than currently available, will be highly beneficial to searching the diffuse disk 
emission for a NT component. 
\medskip

\noindent
{\it Acknowledgement.} 
We acknowledge insightful comments by an anonymous referee that improved clarity and presentation of our work.


\end{document}